\documentclass[aps,prb,twocolumn,showpacs,superscriptaddress,reprint,longbibliography]{revtex4-1}

\usepackage{amsmath, amsfonts, amssymb}
\usepackage{enumerate}
\usepackage{graphicx}
\usepackage{color}
\usepackage[breaklinks]{hyperref}
\usepackage[hyphenbreaks]{breakurl}
\usepackage{paralist}
\usepackage{multirow}

\newcommand{\eref}[1]{Eq.~(\ref{#1})}
\newcommand{\fref}[1]{Fig.~\ref{#1}}

\newcommand{\sref}[1]{Sec.~\ref{#1}}

\newcommand{\e}{\mathrm{e}}

\newcommand{\sx}{\hat{\sigma}_{x}}

\newcommand{\sz}{\hat{\sigma}_{z}}

\newcommand{\ti}{\hat{\tau}_{0}}

\newcommand{\tz}{\hat{\tau}_{z}}

\DeclareMathOperator{\imag}{Im}
\DeclareMathOperator{\sgn}{sgn}

\begin{document}


\title{Subgap states in two dimensional spectroscopy of unconventional superconductors using graphene} 

\newcommand{\aalto}{QTF Centre of Excellence, Department of Applied Physics, Aalto University, 00076 Aalto, Finland}
\newcommand{\bogota}{Departamento de F\'{\i}sica,
	Universidad Nacional de Colombia, Bogot\'a, Colombia}
\newcommand{\bosque}{Departamento de F\'{\i}sica,
	Universidad el Bosque, Bogot\'a, Colombia}
\newcommand{\uam}{Departamento de F\'{\i}sica Te\'orica de la Materia Condensada, Condensed Matter Physics Center (IFIMAC) and Instituto Nicol\'as Cabrera, Universidad Aut\'onoma de Madrid, Spain}
\newcommand{\order}{Preliminary author position}

\author{Oscar Casas Barrera}
\affiliation{\bogota}

\author{Shirley G\'omez P\'aez}
\affiliation{\bogota}
\affiliation{\bosque}

\author{Alfredo Levy Yeyati}
\affiliation{\uam}

\author{Pablo Burset}
\affiliation{\aalto}

\author{William J. Herrera}
\affiliation{\bogota}

\date{\today}


\begin{abstract}
The two-dimensional nature of graphene makes it an ideal platform to explore proximity-induced unconventional planar superconductivity and the possibility of topological superconductivity. 
Using Green's functions techniques, we study the transport properties of a finite size ballistic graphene layer placed between a normal state electrode and a graphene lead with proximity-induced unconventional superconductivity. 
Our microscopic description of such a junction allows us to consider the effect of edge states in the graphene layer and the imperfect coupling to the electrodes. 
The tunnel conductance through the junction and the spectral density of states feature a rich interplay between graphene's edge states, interface bound states formed at the graphene--superconductor junction, Fabry-P\'erot resonances originated from the finite size of the graphene layer, and the characteristic Andreev surface states of unconventional superconductors. Within our analytical formalism, we identify the separate contribution from each of these subgap states to the conductance and density of states. Our results show that graphene provides an advisable tool to determine experimentally the pairing symmetry of proximity-induced unconventional superconductivity. 
\end{abstract}

\maketitle


\section{Introduction}
Unconventional superconductivity involves all pairing states that deviate from the ordinary $s$-wave, spin-singlet Cooper pairs\cite{Sigrist_RMP}, and are thus classified according to the symmetry of their order parameter. 
For example, high-$T_{c}$ superconductors feature an anisotropic $d$-wave spin-singlet pairing state\cite{Kashiwaya_2000,Lee_2006} and there is increasing evidence for the compounds UPt$_{3}$ and Sr$_{2}$RuO$_{4} $ to be spin-triplet chiral $p$-wave superconductors\cite{Mackenzie_2003,*Maeno_2012,*Kallin_2012}. 
Recently, topological superconductors\cite{Read_2000} have triggered an intense research activity as they host gapless Majorana surface states, a candidate for fault-tolerant quantum computing\cite{Ivanov_2001,Alicea_Majorana,*Beenakker_Majorana,*DasSarma_Majorana}. 
Topological superconductivity can be artificially engineered in proximity-induced semiconductor nanowires\cite{Lutchyn_2010,Oreg_2010,Potter_2010,*Potter_2011} or naturally arises on chiral superconductors\cite{Chiral_SC_2016}. 
A chiral superconducting state has also been proposed for other systems, including graphene, where the unconventional superconductivity can come from repulsive interactions\cite{Levitov_2012,Black-Schaffer_2012b,Chiral_graphene_2014} or be induced by proximity to an electron-doped oxide superconductor\cite{Robinson_2017}. 
Tunneling conductance measurements at normal metal--superconductor junctions are a very useful tool to detect signatures of all these types of unconventional superconductivity\cite{Kashiwaya_2000}. 
In a ballistic junction, transport at voltages below the superconducting gap is mediated by Andreev reflections, where incident electrons are converted into holes in the normal metal creating Cooper pairs in the superconductor\cite{BTK,Klapwijk_Review}. 
The presence of surface states in unconventional superconductors is connected to resonance peaks in the Andreev reflection probability, resulting in conductance peaks below the superconducting gap\cite{Tanaka_1995,Kashiwaya_1995,Kashiwaya_1996,Honerkamp_1998,Lu_2016,Burset_2017}. 

\begin{figure}[tbp]
	\includegraphics[width=1.\columnwidth]{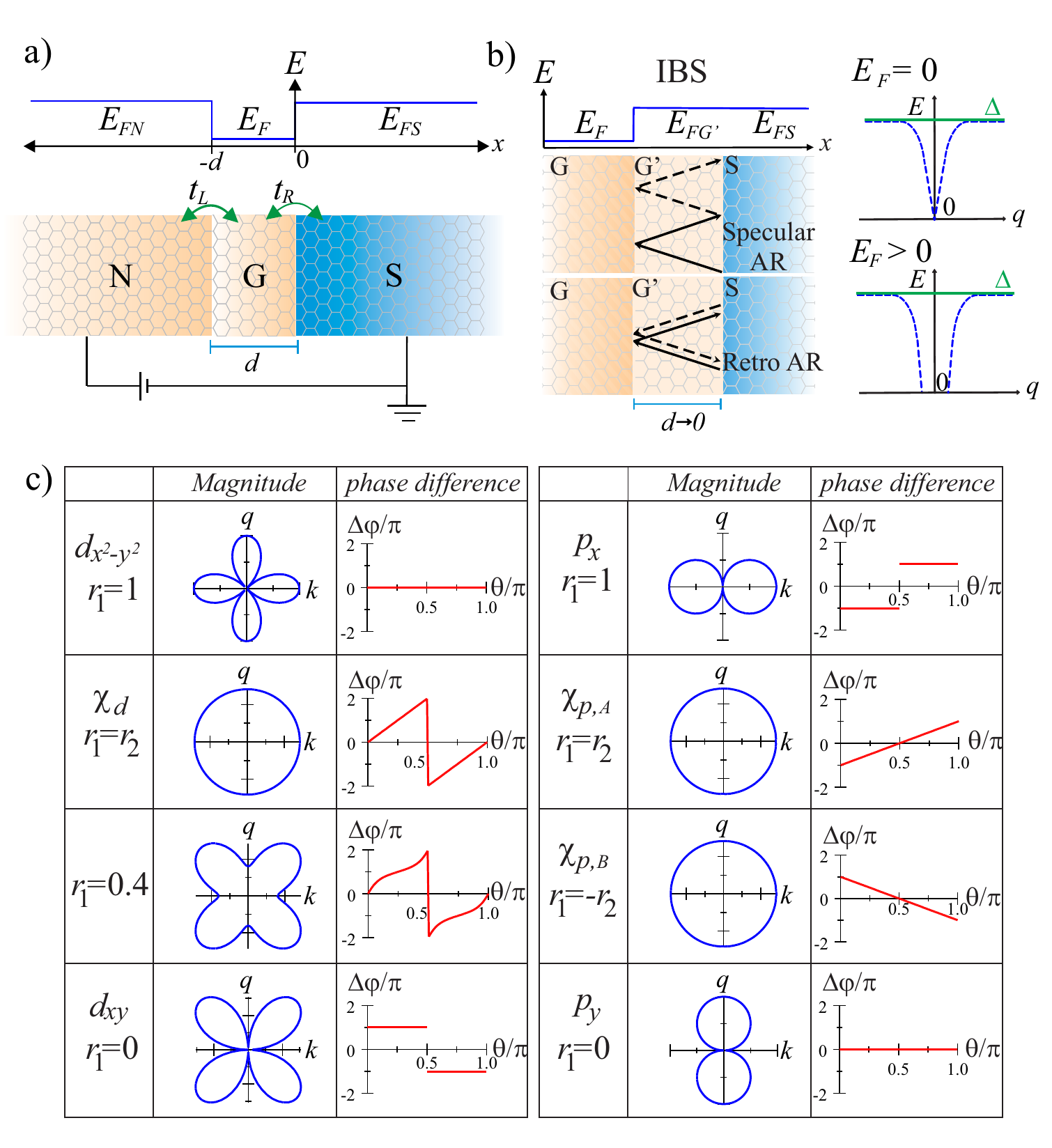}
	\caption{Graphene N-G-S junction. 
	(a) Schematic of the N-G-S junction, including the energy profile. 
	(b) IBS are formed at the G-S junction and can be interpreted as standing waves located at an intermediate graphene layer of width $d\!\rightarrow\!0$ (left panel). 
	Sequences of Andreev specular and retro reflection processes involved in the formation of IBS are sketched: (solid) dashed arrows represent group velocities for (electron-) hole-like quasiparticles. The right panel shows the dispersion relation for the IBS when the superconductor is conventional $s$-wave. 
	(c) Order parameter and angle dependence of the phase difference for the $d$-wave and $p$-wave symmetries considered. See text for more details. 
	}
	\label{fig:sketch}
\end{figure}

Unfortunately, tunneling spectroscopy of subgap resonances presents several experimental challenges, specially for nanoscale devices\cite{Klapwijk_Review}. When considering hybrid junctions where the reservoirs and the intermediate scattering region are built from different materials, as sketched in \fref{fig:sketch}(a), each interface between the intermediate region and the reservoirs may present a different transmission\cite{Jonas_2017}. Additionally, quantum-coherent transport across the junction results in the emergence of Fabry-P\'erot resonances\cite{Tinkham_2001,Kretinin_2010}. All these effects can mask the experimental detection of novel phenomena associated to unconventional superconductivity\cite{Klapwijk_Review}.
However, recent experimental advances involving graphene-based nanoscale devices provide new ways to circumvent these challenges. 

Graphene is a two-dimensional Dirac semimetal with high carrier mobility\cite{Sarma_RMP,Nori_PhysRep,DiracMaterials}. High-quality graphene nanoscale transistors have been achieved\cite{Avouris_2011,*Avouris_2012} and
fabrication of graphene nanoribbons with well-defined edges is an experimental possibility\cite{Liljeroth_2014,*Liljeroth_2015}. Early reports of graphene-based Josephson junctions were assumed to work in the diffusive regime with low-transmitting interfaces\cite{Heersche_2007,*Miao_2007,*Shailos_2007,*Andrei_2008}. 
Recent experiments, however, have achieved good quality ballistic graphene--superconductor contacts\cite{Schonenberger_2012,Klapwijk_2015,Geim_2016}. In particular, encapsulation in hexagonal boron nitride provides high-quality transparent junctions that work in the ballistic regime\cite{Klapwijk_2015,Geim_2016}. Control over the independent doping of the graphene layer has allowed to measure specular Andreev reflections\cite{Efetov_2016}--an unusual type of Andreev process that only manifests when the doping is smaller than the applied voltage and the superconducting gap\cite{Specular}. 
Advances in experimental control of graphene devices are leading to a series of remarkable works reporting spectroscopy of Andreev bound states in Josephson junctions\cite{Dirks_2011}, splitting of Cooper pairs\cite{Hakonen_2015}, and possible proximity-induced superconductivity in graphene, either by growing graphene layers on superconductors\cite{Tonnoir_2013} or by doping it with adatoms\cite{Ludbrook_2015,Chapman_2016}. 
Indeed, the peculiar hexagonal lattice of graphene allows for the formation of unconventional pairing correlations\cite{Uchoa_2007,Jiang_2008,Levitov_2012,Black-Schaffer_2012b}. 
Graphene has been recently grown on top of unconventional (non-chiral) $d$-wave superconductors, revealing an interesting induced $p$-wave pairing state\cite{Robinson_2017}. 
Additionally, a recent experiment reports evidence of intrinsic unconventional superconductivity in graphene superlattices\cite{Jarillo_2018}. 
More experimental and theoretical work is required to fully understand the emergent unconventional superconductivity in graphene and determine if it is chiral and topological. 

In this work, we analyze the transport properties of ballistic junctions consisting of a finite graphene layer contacted by a normal state and a superconducting macroscopic lead [cf. \fref{fig:sketch}(a)]. 
Within our model, we study the most representative two-dimensional unconventional superconductors, including nodal $d$-wave and chiral $p$- and $d$-wave pairing states, and the exotic superconducting states induced by graphene's lattice. 
Our combination of scattering and microscopic Green's function techniques allows us to go beyond previous works in graphene-based superconducting hybrids\cite{Specular,Sengupta_2006,Akhmerov_2007,Tkachov_2007,Linder_2007,Linder_2008,Burset_2008,Black-Schaffer_2008,Burset_2009,Xie_2009,Burset_2012} by including many of the most relevant experimental issues appearing at nanoscale graphene--superconductor junctions. 
Namely, by considering a finite size graphene layer we take into account the Fabry-P\'erot resonances (FPR) present in experiments. 
Additionally, we describe imperfect coupling between the graphene layer and the reservoirs --including the effect of graphene's zigzag edge states (ZZES)-- and analyze the effect of doping the layer close or away from the Dirac point. 
As a result, we present differential conductance calculations with very rich subgap features, where the unconventional surface Andreev bound states (SABS) at the edge of the superconductor are mixed with FPRs, graphene's zigzag edge states, and interface bound states (IBS) formed at the graphene--superconductor junction\cite{Burset_2009}, see \fref{fig:sketch}(b). 
We analytically describe the contribution of each process to the density of states (DOS) and differential conductance. We thus analyze the optimal conditions for the use of graphene to detect signatures of unconventional superconductivity. 

The rest of the paper is organized as follows. In \sref{sec:model} we introduce our model and derive the main formulas for transport observables. We describe the spectral properties of G-S and N-G-S junctions in \sref{sec:spec} and \sref{sec:dos}, respectively. 
Next, in \sref{sec:cond}, we discuss the tunneling spectroscopy of unconventional superconductors. We present our conclusions in \sref{sec:conc}. 
The details of some of the model calculations are given in the Appendix \ref{sec:app1} and \ref{sec:app2}.


\section{Model \label{sec:model}}
Our system consists of a graphene sheet (G) of length $d$ and width $W$ connected to reservoirs as sketched in \fref{fig:sketch}(a).
We consider transport along the $x$-direction and assume that $W\!\gg \!d$, so that there is translational invariance along the graphene--reservoir interfaces. The left and right semi-infinite graphene contacts are in the normal (N) and superconducting (S) states, respectively. 
Low-energy excitations of the coupled system are described by the Dirac-Bogoliubov-de Gennes (DBdG) equations 
\begin{equation}
\begin{pmatrix}
(\hat{H}_{\pm }-\varepsilon _{F}\hat{\sigma}_{0})\hat{s}_{0} & \hat{\Delta}%
\hat{\sigma}_{0} \\ 
\hat{\Delta}^{\dagger }\hat{\sigma}_{0} & (\varepsilon _{F}\hat{\sigma}_{0}-%
\hat{H}_{\pm })\hat{s}_{0}%
\end{pmatrix}%
\begin{pmatrix}
\phi _{e} \\ 
\phi _{h}%
\end{pmatrix}%
\!=\!E%
\begin{pmatrix}
\phi _{e} \\ 
\phi _{h}%
\end{pmatrix}%
,  \label{eq:hamil}
\end{equation}%
with $E\!\geq \!0$ the excitation energy and where the Pauli matrices $\hat{\sigma}_{\nu }$ ($\hat{s}_{\nu }$), with $\nu \!=\!0,1,2,3$, act in lattice (spin) space. The electrostatic potential of each region can be independently fixed and we take $\varepsilon _{F}\!=\!E_{FN},E_{F},E_{FS}$ for regions N, G, and S, respectively. 
In \eref{eq:hamil}, we have decoupled graphene's valley degree of freedom by assuming that the pair potential couples electrons and holes from different valleys\cite{Specular,Beenakker_RMP}. 
For each valley, we use the Dirac Hamiltonian $\hat{H}_{\pm }\!=\!-iv_{F}\partial _{x}\hat{\sigma}_{1}\pm v_{F}q\hat{\sigma}_{2}$, with $v_{F}$ the Fermi velocity and $q$ the conserved component of the wave vector parallel to the interfaces. 
Owing to such valley decoupling, we initially describe $\hat{H}_{+}$ only and later we discuss the role of the other valley. 

The superconducting order parameter is only non-zero in region S, i.e., $\hat{\Delta}(\theta ,x)\!=\!\hat{\Delta}(\theta )\Theta (x)$, with $\Theta (x)$ the Heaviside function, $\theta \!=\!\sin ^{-1}(q/k_{FS})$ the angle in reciprocal space and $k_{F}$ the Fermi wave vector. 
We only consider spin-degenerate unconventional superconductors which allows us to decouple the spin degree of freedom in \eref{eq:hamil}\cite{Sigrist_RMP,Honerkamp_1998,Burset_2014}. 
Indeed, for spin-singlet states, we have $\hat{\Delta}(\theta )\!=\! \Delta(\theta)\mathrm{e}^{i\phi }(i\hat{s}_{2})$, with $\phi$ the global $U(1)$ gauge phase. 
Analogously, for spin-triplet superconductors we take $\hat{\Delta}(\theta )\!=\!\mathrm{e}^{i\phi } \mathbf{d}(\theta)\cdot \hat{\mathbf{s}}(i\hat{s}_{2})$, with the odd vector function $\mathbf{d}(\theta )\!=\!-\mathbf{d}(\pi\!-\!\theta)$. 
As long as the vector $\mathbf{d}$ is perpendicular ($\mathbf{d}\!\propto \!\hat{\mathbf{z}}$) or parallel to the $x-y$ graphene plane, the spin degree of freedom can be decoupled in \eref{eq:hamil}. 
In this equation, the pairing is proportional to the identity matrix in lattice space, $\hat{\sigma}_{0}$, since we consider only on-site induced superconductivity in the graphene lattice\cite{Burset_2008}. 

Under these approximations, the resulting DBdG equations written in Nambu (particle-hole) and lattice spaces read as $\check{H}\psi \!=\!E\psi $. Specifically,  
\begin{equation}
\begin{pmatrix}
v_{F}\mathbf{k}_{\alpha }\cdot \hat{\mathbf{\sigma }}-\mu \hat{\sigma}%
_{0} & \Delta (\theta _{\alpha })\e^{i\phi }\hat{\sigma}_{0} \\ 
\Delta ^{\ast }(\theta _{\alpha })\e^{-i\phi }\hat{\sigma}_{0} & \mu \hat{%
\sigma}_{0}-v_{F}\mathbf{k}_{\alpha }\cdot \hat{\mathbf{\sigma }}%
\end{pmatrix}%
\begin{pmatrix}
\phi _{\alpha }^{e} \\ 
\phi _{\alpha }^{h}%
\end{pmatrix}%
\!=\!E%
\begin{pmatrix}
\phi _{\alpha }^{e} \\ 
\phi _{\alpha }^{h}%
\end{pmatrix}%
,  \label{eq:hamil_red}
\end{equation}%
with $\alpha \!=\!\pm $ for right and left movers, respectively. 
We notice here that, since the lattice is acting as a  pseudo-spin degree of freedom, the reduced Hamiltonian of \eref{eq:hamil_red} is also suitable to describe induced pairing amplitudes with an structure in lattice space\cite{Uchoa_2007,Black-Schaffer_2007,Jiang_2008,Robinson_2017}, with the appropriate redefinition of the pair potential. 

To take into account the sign change of the triplet state with the wave vector, we only consider $k_{x}\!\geq \!0$ and define $\mathbf{k}_{\alpha}\!=\!(\alpha k_{x},q)$. We thus set $\theta _{+}\!=\!\theta $ and $\theta
_{-}\!=\!\pi \!-\!\theta $ for right and left movers, respectively. 
The pair potential adopts the general form 
\begin{equation}
\Delta (\theta_{\pm} )=\Delta _{0}\left[ r_{1}\cos (n\theta_{\pm}) +ir_{2}\sin( n\theta_{\pm})
\right] ,  \label{eq:pairing}
\end{equation}%
with $\Delta _{0}\!\geq \!0$ the potential amplitude and $r_{2}^{2}\!=\!1\!-\!r_{1}^{2}$ the relative value of real and imaginary parts of the pair potential. 
The integer $n\!=\!0,1,2,\dots $ determines the orbital symmetry of the pairing state, i.e., the values $n\!=\!0,2$ correspond $s$- and $d$-wave
states, respectively, while $n\!=\!1$ represents a $p$-wave state. 
More details about \eref{eq:hamil_red} and its solutions can be found in the Appendix \ref{sec:app1}.

The retarded/advanced Green functions associated to the Hamiltonian in \eref{eq:hamil_red} satisfies the non-homogeneous DBdG equation
\begin{equation}
\left[ \check{H}-(E\pm i0^{+})\check{I}\right] \check{g}_{q}^{r,a}(x,x^{\prime })=\delta \left(
x-x^{\prime }\right) \check{I}\text{,} 
\label{eq:Green_equation}
\end{equation}
where $\check{g}_{q}^{r,a}(x,x^{\prime })$ is the Fourier transform of the spatial Green function on the coordinates parallel to the interfaces, $\check{I}$ is the four-dimensional identity matrix and $\check{H}$ is the DBdG Hamiltonian given by \eref{eq:hamil_red}. 
The unperturbed Green's function $\check{g}_{q}^{r,a}(x,x^{\prime })$ is obtained combining asymptotic solutions that obey boundary conditions at the edges of a finite length graphene sheet, following a generalization of the method developed in Refs.~\onlinecite{McMillan_1968,Furusaki_1991,Kashiwaya_2000,Herrera_2010,Burset_2015,Crepin_2015,Breunig_2018} for unconventional superconductors and described in Appendix \ref{sec:app2}. 

The Green functions of the coupled system $\check{G}$ are calculated by means of an algebraic Dyson equation of the form\cite{Herrera_2010,Gomez_2011} 
\begin{equation}
\check{G}_{q,ij}^{r,a}=\check{g}_{q,ij}^{r,a}+\check{g}_{q,ik}^{r,a}\check{\Sigma}_{kl}\check{G}_{q,lj}^{r,a}%
\text{,} 
\label{eq:Dyson_equation}
\end{equation}
with short-hand notation $\check{g}_{q,ij}^{r,a} \!=\! \check{g}_{q}^{r,a}(x_{i},x_{j}^{\prime})$. The self-energies $\check{\Sigma}_{kl}$ ($k\neq l$) represent the hopping matrix between two different regions\cite{Herrera_2010}. 
For the zigzag boundary conditions adopted in this
work, opposite edges of the graphene layer correspond to atoms from a different sublattice. This leads to a specific form of the hopping matrix as defined below. 

\subsection{Transport observables}
The spectral density of states is calculated from the retarded Green
function as 
\begin{equation}  \label{eq:spec-den}
\mathcal{A}_{q}(x,E)=-\frac{1}{\pi }\imag\left\{ \mathrm{Tr}\check{G}%
_{q,ee}^{r}(x,x,E)\right\} ,
\end{equation}%
where the trace is taken
over the electron-electron component in Nambu space. The local density of
states (DOS) is given by 
\begin{equation}
\rho (x,E)=\sum_{q,\sigma}\mathcal{A}_{q}(x,E).  \label{eq:dos-def}
\end{equation}

The current for the setup sketched in Fig.~\ref{fig:sketch}(a) is obtained
following the Hamiltonian approach\cite{Cuevas_1996,Gomez_2011}. We compute the charge tunneling between the regions from the tight-binding Hamiltonian 
\begin{equation}
\hat{H}=\hat{H}_{L}+\hat{H}_{C}+\hat{H}_{R}+\hat{H}_{T_{L}}+\hat{H}_{T_{R}}%
\text{,}
\end{equation}%
where $\hat{H}_{i=L,C,R}$ are the unperturbed Hamiltonians, and $\hat{H}%
_{T_{L,R}}$ is the tunneling Hamiltonian of the form 
\begin{align}
\hat{H}_{T_{L}}& =\sum_{q,\sigma }t_{L}c_{q,LA\sigma }^{\dagger }\psi
_{q,B\sigma }(-d)+\text{h.c.}, \\
\hat{H}_{T_{R}}& =\sum_{q,\sigma }t_{R}c_{q,RB\sigma }^{\dagger }\psi
_{q,A\sigma }(0)+\text{h.c.},
\end{align}%
where $c_{q,\nu j\sigma }$ and $\psi _{q,j}(x)$, with $\nu =L,R$ and $j=A,B$%
, are annihilation for electrons at the edges of the $L$,$R$,$C$ regions
with parallel momentum $q$. The average current through the left interface is given by 
\begin{gather}
I=-e\left\langle \frac{d}{d\tau }\hat{N}_{L}\right\rangle  \\
=\frac{ie}{h}t_{L}\sum\limits_{q,\sigma }\left( \left\langle \hat{c}%
_{q,LA\sigma }^{\dag }\hat{\psi}_{q,B\sigma }(-d)\right\rangle -\left\langle 
\hat{\psi}_{q,B\sigma }^{\dagger }(-d)\hat{c}_{q,LA\sigma }\right\rangle
\right) .  \notag
\end{gather}%
This average can be expressed in terms of the Keldysh or non-equilibrium
Green functions defined as 
\begin{equation}
\check{G}_{q,ij}^{\alpha \beta }\left( \tau _{\alpha },\tau _{\beta
}^{\prime }\right) =-i\left\langle \hat{T}\left[ \hat{D}_{q,i}\left( \tau
_{\alpha }\right) \hat{D}_{q,j}^{\dag }\left( \tau _{\beta }^{\prime
}\right) \right] \right\rangle \text{,}
\end{equation}%
\begin{equation*}
\hat{D}_{q,i}^{\dag }\left( \tau \right) =\left( \hat{d}_{q,iA\uparrow
}^{\dag }\left( \tau \right) ,\hat{d}_{q,iB\uparrow }^{\dag }\left( \tau
\right) ,\hat{d}_{q,iA\downarrow }\left( \tau \right) ,\hat{d}%
_{q,iB\downarrow }\left( \tau \right) \right) ,
\end{equation*}%
where $i,j=L,C,C^{\prime },R$, $\ \hat{d}_{q,Lj\sigma }(\tau )=\hat{c}%
_{q,Lj\sigma }(\tau )$, $\hat{d}_{q,Cj\sigma }(\tau )=\hat{\psi}_{q,j\sigma
}(-d,\tau )$, $\hat{d}_{q,C^{\prime }j\sigma }(\tau )=\hat{\psi}_{q,j\sigma
}(0,\tau )$, $\hat{d}_{q,Rj\sigma }(\tau )=\hat{c}_{q,Rj\sigma }(\tau )$,
superscripts $\alpha ,\beta $ correspond to the temporal branches of the
Keldysh contour and $\hat{T}$ is the Keldysh time-ordering operator. Then,
the current evaluated at the left juncture reduces to 
\begin{equation*}
I=\frac{e}{h}\sum\limits_{q,\sigma }\int \mathrm{d}E\mathrm{Tr}\left( \check{%
	\tau}_{z}\check{t}_{L}\left[ \check{g}_{q,L}^{+-}\check{t}%
_{L}^{\dag }\check{G}_{q,CC}^{-+}-\check{g}_{q,L}^{-+}\check{t}%
_{L}^{\dag }\check{G}_{q,CC}^{+-}\right] \right) \text{,}
\end{equation*}%
where $\check{t}_{L,R}=\check{\Sigma}_{LC,C'R}=(t_{L,R}/2\hat{\tau}_{z})(\hat{\sigma}_{x}+i\hat{%
	\sigma}_{y})$ are the coupling self-energies, $\check{g}_{q,L}^{\gamma
	=+-,-+}$ corresponds to the non-equilibrium Green functions for the left
uncoupled electrode evaluated at $x=x^{\prime }=-d$, and $\check{G}%
_{q,CC}^{\gamma }$ is the non-equilibrium Green function of the coupled
system evaluated at $x=x^{\prime }=-d+0^{+}$. The last expression can be
written in terms of the retarded/advanced Green functions by means of the
following Dyson equation that contains information of the full region at the
right of the left juncture 
\begin{eqnarray}
\check{G}_{q,CC}^{\gamma } &=&\check{G}_{q,CC}^{r}\check{t}_{L}\check{g}%
_{q,L}^{\gamma }\check{t}_{L}^{\dag }\check{G}_{q,CC}^{a}+ \\
&&\check{G}_{q,CC^{\prime }}^{r}\check{t}_{R}^{\dag }\check{g}%
_{q,R}^{\gamma }\check{t}_{R}\check{G}_{q,C^{\prime }C}^{a},  \notag
\end{eqnarray}%
with $\check{g}_{q,i}^{+-}\!=\!2\pi i\check{\rho}_{q,i}\check{f}_{i}$, $%
\check{g}_{i}^{-+}\!=\!-2\pi i\check{\rho}_{q,i}\left( \hat{\tau}_{0}\hat{%
	\sigma}_{0}-\check{f}_{i}\right) $ and $\check{\rho}_{q,i}\!=\!\mp \imag(%
\check{g}_{q,i}^{r(a)})/\pi $. The Fermi-Dirac distribution matrix for a
voltage $V_{i}$ applied to electrode $i$ is defined as $\check{f}%
_{i}(\varepsilon ,V)\!=\!\mathrm{diag}(f(\varepsilon \!-\!eV_{i})\hat{\sigma}%
_{0},f(\varepsilon \!+\!eV_{i})\hat{\sigma}_{0})$, with $f({\varepsilon }%
)\!=\![1\!+\!\exp {(\beta \varepsilon )}]^{-1}$ and $\beta $ the inverse
temperature. Finally, the current is given by 
\begin{gather*}
I=\sum_{q}I_{q}=\frac{4\pi ^{2}e}{h}\sum\limits_{q,\sigma }\int \mathrm{d}E%
\mathrm{Tr}\left\{ \tau _{z}\check{t}_{L}\check{\rho}_{q,L}\left(
{}\right. \right.  \\
\left[ \check{f}_{L}\check{t}_{L}^{\dag }\check{G}_{q,CC}^{r}\check{%
	t}_{L}\check{\rho}_{q,L}-\check{t}_{L}^{\dag }\check{G}_{q,CC}^{r}%
\check{t}_{L}\check{\rho}_{q,L}\check{f}_{L}\right] \check{t}%
_{L}^{\dag }\check{G}_{q,CC}^{a}+ \\
\left. \left. \left[ \check{f}_{L}\check{t}_{L}^{\dag }\check{G}%
_{q,CC^{\prime }}^{r}\check{t}_{R}^{\dag }\check{\rho}_{q,R}-\check{%
	t}_{L}^{\dag }\check{G}_{q,CC^{\prime }}^{r}\check{t}_{R}^{\dag }%
\check{\rho}_{q,R}\check{f}_{R}\right] \check{t}_{R}\check{G}%
_{q,C^{\prime }C}^{a}\right) \right\} .
\end{gather*}%
The differential conductance as a function of $q$ reads 
\begin{equation}
\sigma _{q}=\frac{\partial I_{q}}{\partial V}=\sigma _{q,A}+\sigma _{q,Q},
\label{eq:cond}
\end{equation}%
where $\sigma _{q,A}$ is the contribution of Andreev processes, 
\begin{equation}
\sigma _{q,A}=\frac{16e^{2}}{h}\mathrm{Tr}\left[ \mathrm{Re}\left\{ \bar{\rho%
}_{q,Lee}\hat{G}_{q,CCeh}^{r}\bar{\rho}_{q,Lhh}\hat{G}_{q,CChe}^{a}\right\} %
\right] ,  \label{eq:cond_andreev}
\end{equation}%
with $\bar{\rho}_{q,i\mu \nu }$ the Nambu component of the matrix $\bar{\rho}%
_{q,i}\!=\!\pi \check{t}_{i}\check{\rho}_{q,i}\check{t}%
_{i}^{\dagger }$. The contribution $\sigma _{q,Q}$ due to quasiparticles is given by 
\begin{gather}
\sigma _{q,Q}=\frac{8e^{2}}{h}\mathrm{Tr}\left[ \mathrm{Re}\left\{ \bar{\rho}%
_{q,Lee}\left( {}\right. \right. \right.   \label{eq:cond_qp} \\
\left. \left. \hat{G}_{q,CC^{\prime }ee}^{r}\left[ \bar{\rho}_{q,Ree}\hat{G}%
_{q,C^{\prime }Cee}^{a}-\bar{\rho}_{q,Reh}\hat{G}_{q,C^{\prime
	}Che}^{a}\right] \right. \right.   \notag \\
-\left. \left. \left. \hat{G}_{q,CC^{\prime }eh}^{r}\left[ \bar{\rho}%
_{q,Rhe}\hat{G}_{q,C^{\prime }Cee}^{a}-\bar{\rho}_{q,Rhh}\hat{G}%
_{q,C^{\prime }Che}^{a}\right] \right) \right\} \right] .  \notag
\end{gather}%
We use a highly doped semi-infinite
graphene lead in order to model the normal electrode. Conductances are
normalized to the normal-state graphene conductance. Eq.~(\ref{eq:cond}) provides
a generalized formula to calculate the differential conductance in
graphene-superconductor hybrid structures.

\begin{figure*}
	\includegraphics[width=1.\textwidth]{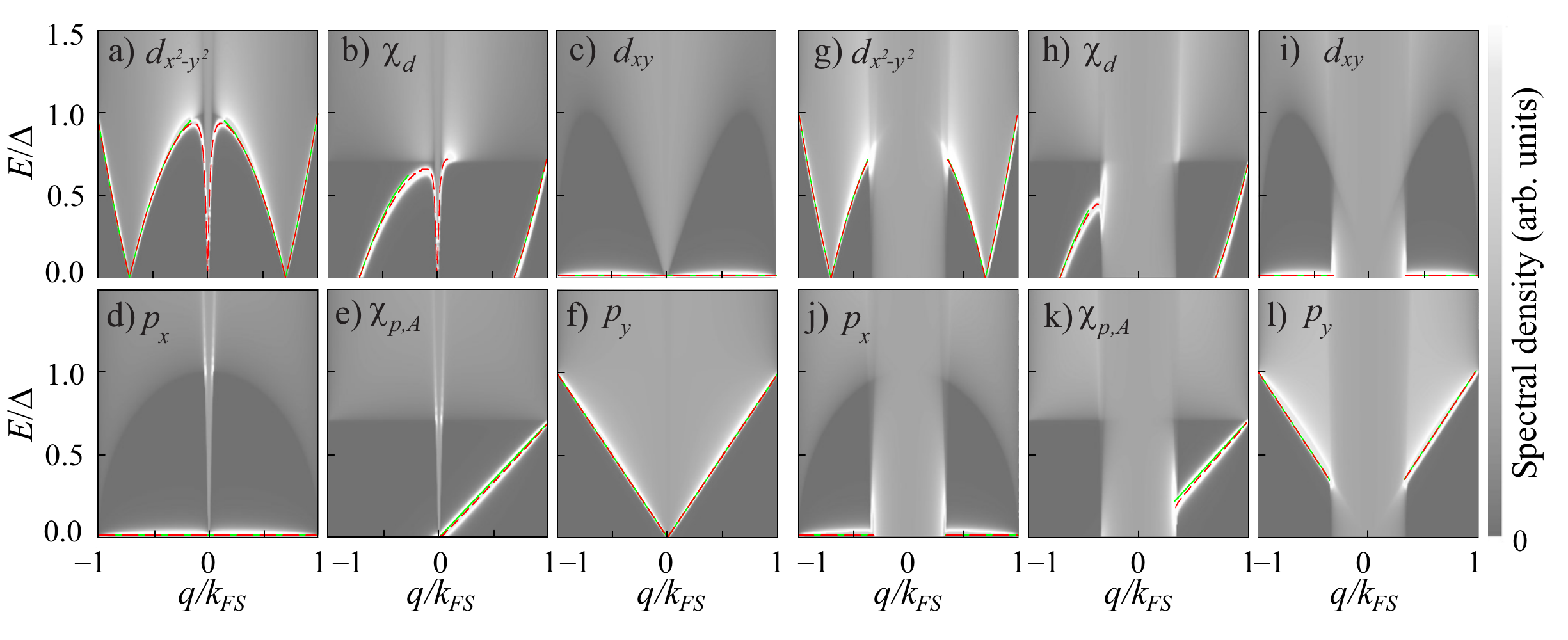}
	\caption{Spectral density of states of a semi-infinite
		graphene sheet coupled to a semi-infinite anisotropic superconductor evaluated at the interface. $E_{F}=0$ for figures (a-f) and $E_{F}=10\Delta$ for (g-l). SABS (solid green lines) and IBS (dashed red lines) are depicted for the different symmetries. }
	\label{fig:IBS}
\end{figure*}

\section{Spectral properties of graphene-superconductor junctions \label{sec:spec}}
The spectral properties of the full N-G-S system contain information from many different sources: ZZES, FPR, IBS and SABS. We analyze the N-G-S setup numerically in the next section. 
In some particular cases, however, we can obtain simple analytical formulas for the contribution from some of these states. In this section, we consider a simpler setup where the left electrode is removed, resulting in a graphene-superconductor (G-S) junction, see \fref{fig:sketch}(b). 
The coupled Green function is given by Dyson's equation introduced in \eref{eq:Dyson_equation}, see Appendix \ref{sec:app2} for more details. 
The denominator of this perturbed Green function encodes information about the different states present in the junction. By finding its zeros, we obtain the dispersion relation of the induced resonances in the junction. 
In this section, we assume a perfect coupling between the graphene layer and the superconductor to avoid the formation of ZZES at the G-S interface. However, for finite graphene layers there is another ZZES at the opposite edge. 
Analyzing the Green function at the middle of the graphene layer, we can minimize the impact of ZZES on the spectrum and focus on the other resonances. 

First, proximity-induced pairing from an unconventional superconductor always manifests with the emergence of SABS with a dispersion relation given by\cite{Lofwander_2001}
\begin{equation}
E_\text{SABS}=\pm \left\vert \Delta \left( \theta \right) \right\vert \cos \Delta
\varphi /2, \label{eq:SABS}
\end{equation}
with $\Delta \varphi \!=\! \varphi _{+}-\varphi _{-}$ the phase difference between the pair potentials $\Delta(\theta_{+})$ and $\Delta(\theta_{-})$ defined in \eref{eq:pairing}. 
\eref{eq:SABS} takes different forms depending on the symmetry of the pair potential, cf. \fref{fig:sketch}(c). 
For symmetries $s,d_{x^{2}-y^{2}}$ and $p_{y}$ ($\Delta \varphi \!=\!0$) we get $E\!=\!\pm \left\vert
\Delta \left( \theta \right) \right\vert $. 
For $d_{xy}$ and $p_{x}$ symmetries, we have $\Delta \varphi \!=\!\pi $ which corresponds to zero energy states (ZES) $E\!=\!0$. 
Chiral symmetries feature an angle dependence as follows:  chiral $d$-wave symmetry $\chi_{d}$ ($\Delta \varphi \!=\! 4\theta $) results in $E\!=\!\pm \left\vert \Delta
\left( \theta \right) \right\vert \cos 2\theta $, and chiral $p$-wave $\chi_{p}$ ($\Delta \varphi \!=\! \mathrm{sgn}(r_{1})2\theta +\pi $) gives $E\!=\! \pm
\left\vert \Delta \left( \theta \right) \right\vert \sin \theta $. 

In addition to the SABS, the special properties of the graphene layer lead to the emergence of IBS\cite{Burset_2009}, which describe how the linear dispersion of graphene adapts to the presence of the gapped superconducting density of states. 
Moreover, a finite graphene layer will feature discrete energy bands, labeled here FPR, and ZZES. 
A general expression that describes IBS and FPR reads as
\begin{gather}
E_\text{G-S}= \label{eq:IBS-FP} \\ \pm \left( \frac{C\cos \left( \Delta \varphi /2\right) +i\sin
\left( \Delta \varphi /2\right) }{\sqrt{C^{2}-1}}\right) \left\vert \Delta
\left( \theta \right) \right\vert , \nonumber 
\end{gather}
with
\begin{equation*}
C =\frac{1+\e^{-i\alpha _{e}}\e^{i\alpha _{h}}h_{+}h_{-}}{\e^{-i\alpha
_{e}}h_{+}+\e^{i\alpha _{h}}h_{-}} , \,
h_{\pm } =\frac{\e^{\pm 2idk_{e(h)}}-1}{\e^{\mp 2i\alpha _{\pm }}\e^{\pm 2idk_{e(h)}}+1} .
\end{equation*}
Here, $\e^{i\alpha_{e(h)}}$ is associated with the angle of incidence of quasiparticles in the graphene region and is defined in Appendix \ref{sec:app2}. For $s$-wave \eref{eq:IBS-FP} coincides with the results in Ref.~\onlinecite{Burset_2009}. For this symmetry, the dispersion relation of the IBS tends to zero at $q\!\rightarrow\!0$ and approaches asymptotically the superconducting gap for large $q$ as it is sketched in \fref{fig:sketch}(b). The IBS are localized at the G-S interface for $E_F\!\gg\!\Delta$ (retro-reflection regime) but can decay over long distances inside the graphene region when $E_F\!\ll\!\Delta$ ( specular reflection regime). 

We now consider two specific cases where \eref{eq:IBS-FP} can be simplified to isolate the contribution from either IBS or FPR (both in the presence of SABS). 

\subsection{Low-doped semi-infinite graphene layer: Interface bound states}
By considering now a semi-infinite graphene layer coupled to a superconductor, we can ignore the geometrical FPR and ZZES and focus on the dispersion relation corresponding to the SABS and IBS. 
By coupling transparently the semi-infinite graphene Green function to the superconducting electrode, we obtain a dispersion relation for the IBS that corresponds to taking the limit $d\!\rightarrow\!\infty$ in \eref{eq:IBS-FP}, where $h_{\pm}\!\rightarrow\!-1$. 
In the heavily-doped regime with $E_F\!\gg\!E,\Delta$, the dispersion relation is given by \eref{eq:SABS} --the IBSs only appear for low doping levels comparable to the superconducting gap. 

At the opposite limit, i.e., close to the Dirac point, $E_{F}\!=\!0$, we find that \eref{eq:IBS-FP} yields
\begin{equation}\label{eq:IBS_EF0} 
\begin{gathered}
E_\text{IBS} \approx \pm \varepsilon \left\vert \Delta \left( \theta \right) \right\vert \\
\times \sqrt{\frac{\left( 1-\eta ^{2}\right) \left( \varepsilon ^{2} + \left\vert\Delta \left( \theta \right) \right\vert ^{2}  \right) }{\varepsilon^{4} + \left\vert \Delta \left( \theta \right) \right\vert ^{4} - 2\varepsilon^{2}\left\vert \Delta \left( \theta \right) \right\vert ^{2}\left( 2\eta^{2}-1\right) } } , 
\end{gathered}
\end{equation}
with $\varepsilon \!=\!\hbar v_Fq$ and $\eta \!=\!\sin \left( \Delta \varphi/2\right) $. 

We show the spectral density of states, \eref{eq:spec-den}, for a semi-infinite graphene layer coupled to a superconductor in \fref{fig:IBS}. We evaluate \eref{eq:spec-den} at the graphene-superconductor interface and consider different pairing symmetries according to \fref{fig:sketch}(c). All results are evaluated for one of graphene's valleys and can show an asymmetry in the momentum $\hbar q$. This asymmetry is explained in detail in the next section. 
The continuous band is shown in gray, with the subgap resonances appearing in bright over the dark background of the superconducting gap. 
The resonances have been fitted using the formulas derived in this section, solid green lines for \eref{eq:SABS} (SABS) and dashed red lines for \eref{eq:IBS-FP} (IBS with $E_F\!\neq\!0$) and \eref{eq:IBS_EF0} (IBS with $E_F\!=\!0$). 
The close similarity between red and green lines demonstrates how SABS and IBS are connected in the semi-infinite layer. This setup corresponds to an ideal case where induced pairing in graphene is mostly given by the unconventional pair amplitude in the superconductor, without spurious effects from FPR or ZZES. 
Indeed, for symmetries $s$, $d_{x^{2}-y^{2}}$ and $p_{y}$ the dispersion relation in \eref{eq:IBS_EF0} reduces to $E\!=\!\pm \varepsilon \left\vert \Delta \left( \theta \right) \right\vert [\varepsilon^{2}\!+\!\left\vert \Delta \left( \theta \right) \right\vert ^{2}]^{-1/2}$ as can be seen in the red dashed plot of \fref{fig:IBS}($a$) and ($f$). For $d_{xy}$ and $p_{x}$ symmetries, \eref{eq:IBS_EF0} features a ZES, see \fref{fig:IBS}($c$) and ($d$). 
For chiral symmetries, $\eta\!=\!\mathrm{sgn}(r_{1})\sin \left( \Delta \varphi /2\right) $ and $\left\vert \Delta \left( \theta \right) \right\vert \!=\!\Delta /\sqrt{2}$ in \eref{eq:IBS_EF0} [\fref{fig:IBS}(b) and (e)]. 

\subsection{Heavily-doped finite graphene layer: Fabry-P\'erot resonances}
A finite graphene layer develops FPR and ZZES. 
By taking the heavily-doped limit, we get rid of the effect of both ZZES and IBS, leaving only the interplay between the geometric FPR and the proximity-induced SABS. 
The dispersion relation \eref{eq:IBS-FP} takes the simple form
\begin{equation}
E_\text{FPR}=\pm \left\vert \Delta \left( \theta \right) \right\vert \cos \left(
\left( k_{e}-k_{h}\right) d-\frac{\Delta \varphi }{2}\right) \text{.}
\label{eq:FP}
\end{equation}
where $k_e$ and $k_h$ are defined in the Appendix. Following Ref.~\onlinecite{Lofwander_2001}, we approximate $k_{e}-k_{h}\simeq 2E/\hbar v_F$ to obtain 
\begin{equation}
E=\frac{\hbar v_F}{2d}\left( \pm 2\pi n+\frac{\Delta \varphi }{2} + \cos ^{-1}\left( \pm E/\left\vert \Delta \left( \theta \right) \right\vert \right)  \right) .
\end{equation}
Note that the separation between energy levels decreases with the length $d$ of the stripe and, therefore, the number of levels per unit of energy --and thus the number of conductance peaks-- increases with $d$. 
\eref{eq:FP} can be interpreted as the intersection points between a straight line with slope $\left\vert \Delta \left( \theta \right) \right\vert ^{-1}$ and a harmonic function with frequency $2d/\hbar v_F$ and phase $\Delta \varphi /2$. 
For pairing symmetries with a dependence on the angle of incidence $\theta$, as $\Delta \varphi $ continuously changes from $0$ to $\pi$, the harmonic function in \eref{eq:FP} shifts from $\cos x$ to $\sin x$ and the intersection points approach to $E\!=\!0$. This induces a gradual shifting of the crests to the center and finally the emergence of a ZES state. As a result, there is a shifting of resonance peaks in the differential conductance $\sigma $ until the appearance of a ZBCP (more details in \sref{sec:cond}). 

\begin{figure*}
	\includegraphics[width=1.\textwidth]{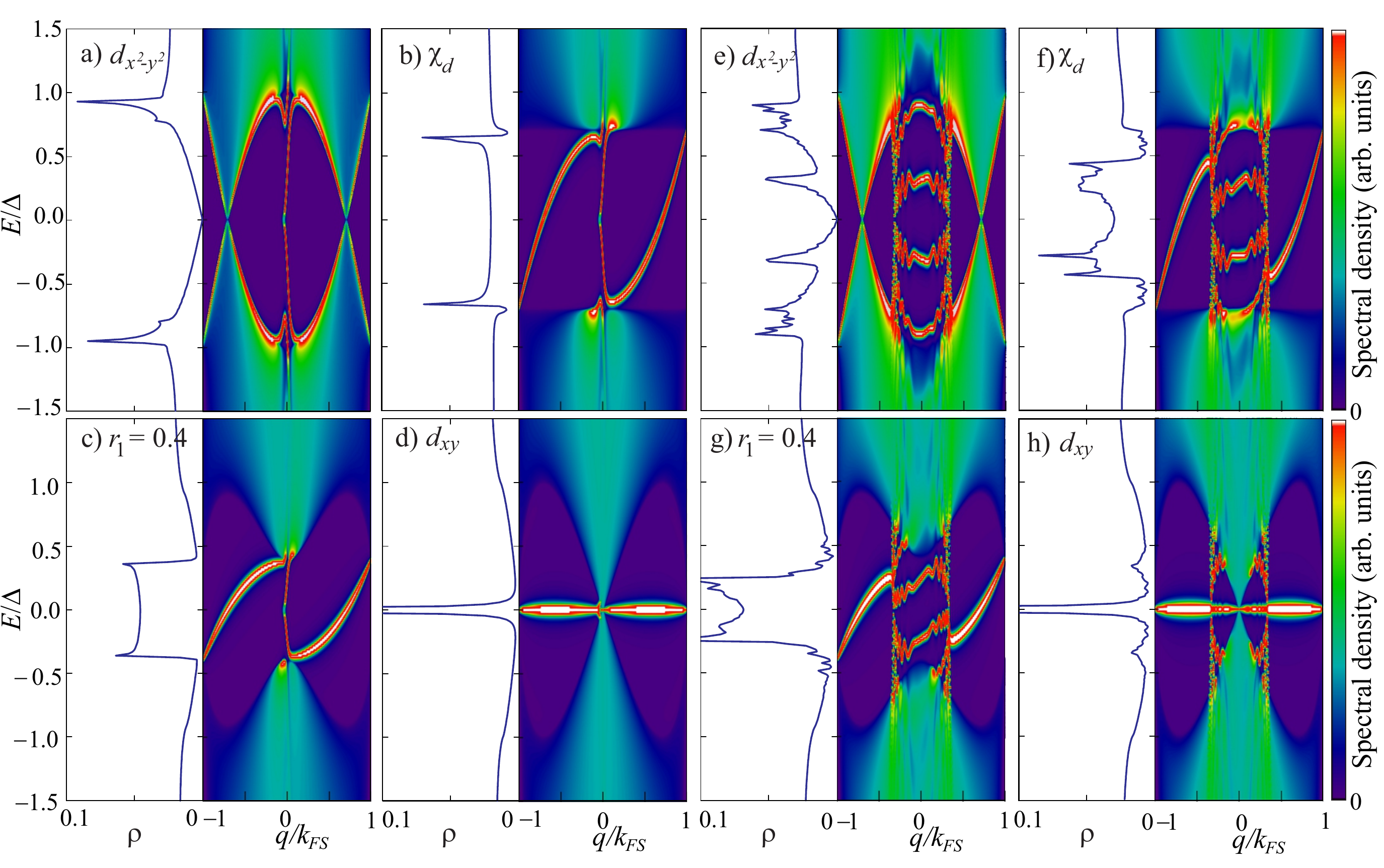}
	\caption{Spectral density and DOS of the graphene N-G-S
		junction at $x=0$. Results for $d$-wave superconductors with different values of the pairing phase. The Fermi energy of the graphene layer is $E_{F}\!=\!0\Delta $ for (a,b,c,d) and $E_{F}\!=\!10\Delta $ for (e,f,g,h). In all cases, $E_{FN}\!=\!E_{FS}\!=\!30\Delta $, $t_{L}\!=\!0.1$, $t_{R}\!=\!1$, and $d\!=\!2\protect\xi$. 
		}
	\label{fig:dos}
\end{figure*}


\begin{figure*}
	\includegraphics[width=1.\textwidth]{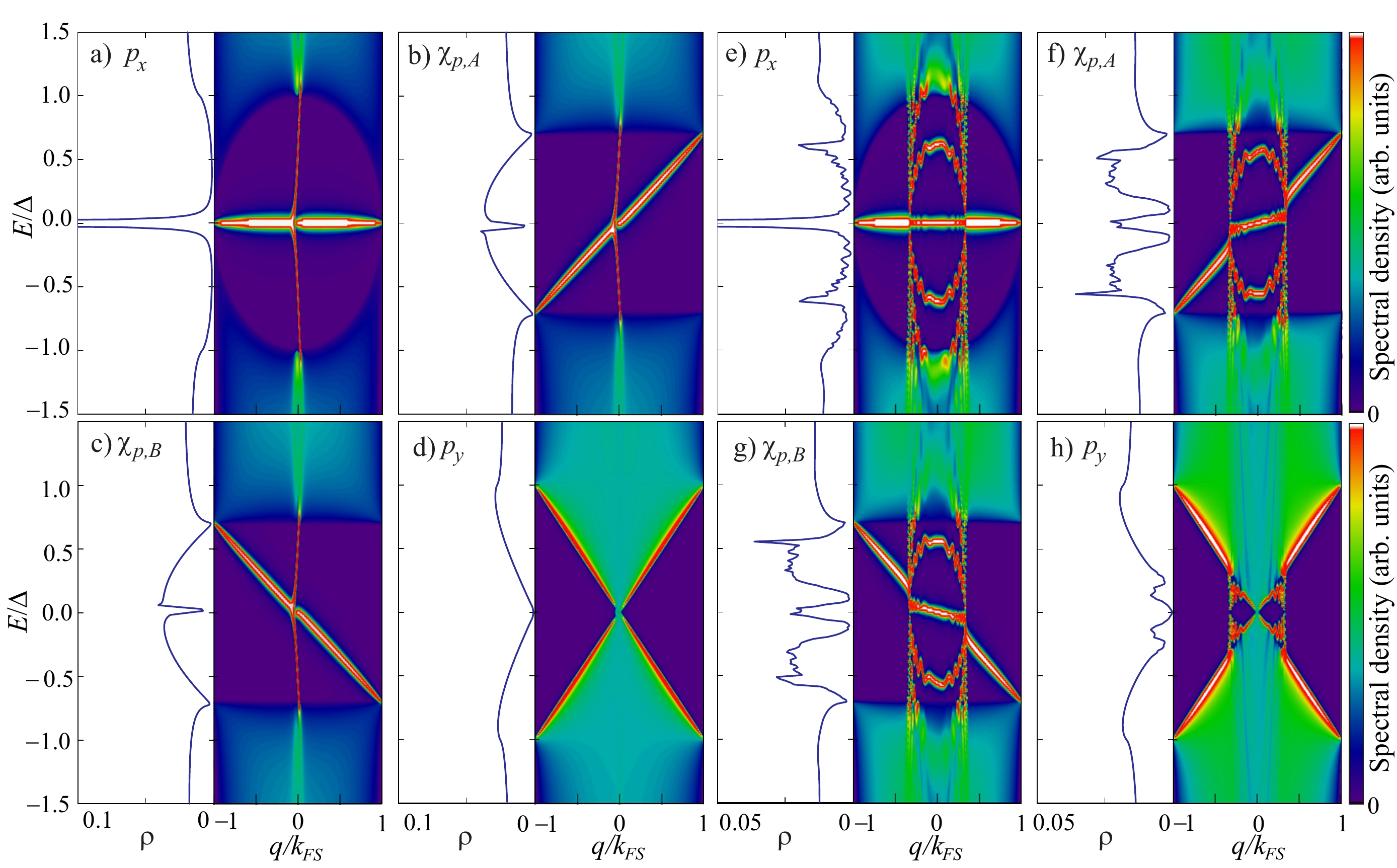}
	\caption{Spectral density and DOS of the graphene N-G-S
		junction at $x=0$. Results for different types of $p$-wave superconductors. The Fermi energy of the graphene layer is $E_{F}\!=\!0\Delta $ for (a,b,c,d) and $E_{F}\!=\!10\Delta $ for (e,f,g,h). In all cases, $E_{FN}\!=\!E_{FS}\!=\!30\Delta $, $t_{L}\!=\!0.1$, $t_{R}\!=\!1$, and $d\!=\!2\protect\xi$. 
		}
	\label{fig:tres}
\end{figure*}

\begin{figure}
	\includegraphics[width=0.95\columnwidth]{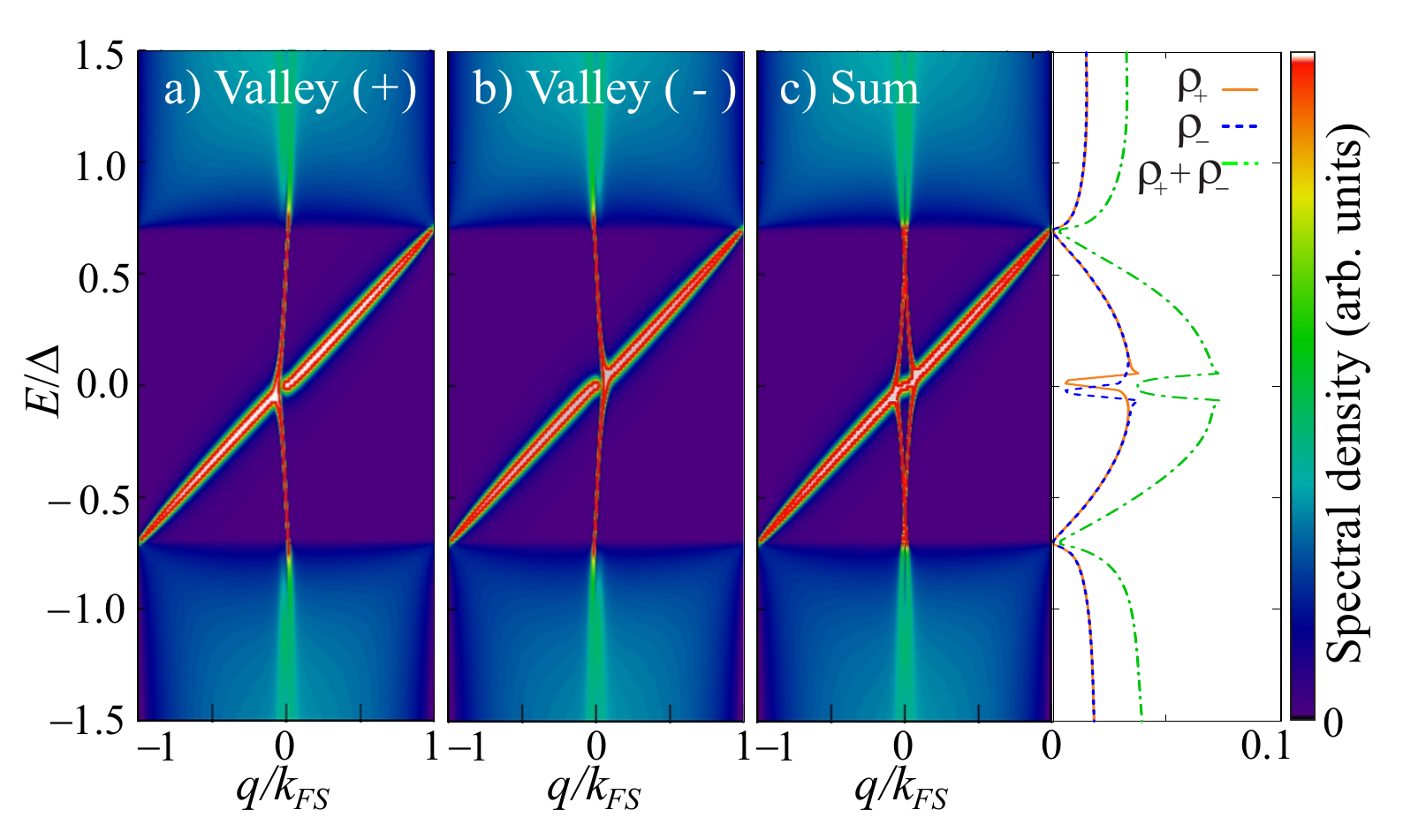}
	\caption{Valley dependence of the spectral density and DOS of the graphene N-G-S junction at $x=0$. All plots are calculated for the chiral $p$-wave symmetry. }
	\label{fig:dos_valley}
\end{figure}

\section{Density of states of the N-G-S junction \label{sec:dos}}
We now focus on the N-G-S junction sketched in \fref{fig:sketch}(a). 
The intermediate region is a graphene layer with zigzag edges along the $y$-direction. When uncoupled from the reservoirs, the isolated graphene layer features localized zigzag edge states (ZZES) at $E\!=-E_{F}$. The coupling to the leads, controlled by the interface transparencies $t_{L,R}$, splits the ZZES which completely vanish at perfect transparency\cite{Brey_2006,Tkachov_2007,Burset_2009,Tkachov_2009,Tkachov_2009b,Herrera_2010}. The ZZES have a resonant contribution to the density of states, with magnitude much bigger than that of any other resonances or Andreev states considered here. They thus play an important role in the tunneling properties, as shown in the next section. 
In what follows, we take $t_{L}\!=\!0.1$ and $t_{R}\!=\!1$. 
A perfect coupling to the superconductor ($t_{R}\!=\!1$) guarantees that at the interface there is no ZZES and only Andreev reflections take place. 
By considering the coupling to the N lead in the tunnel regime ($t_{L}\!=\!0.1$), we include an important contribution from normal backscattering processes and from the ZZES at that edge. We choose the width of the graphene layer to be $d\!=\!2\xi$, with $\xi\!=\!\hbar v_F/\Delta$ the superconducting coherence length. This is enough for Andreev processes to contribute to the conductance to the leftmost electrode, but it also reduces the effect of the ZZES on the density of states calculated close to the G-S interface. Additionally, the reduced N-G tunneling increases the finite-size effects at the intermediate region. 

Under such conditions, we plot in \fref{fig:dos} the spectral density and DOS, calculated, respectively, from \eref{eq:spec-den} and \eref{eq:dos-def} close to the G-S interface ($x\!\rightarrow\!0$), for the $d$-wave symmetries. An equivalent plot for $p$-wave symmetries is shown in \fref{fig:tres}. 
Before analyzing the effect of the different pairing symmetries, we first comment on some common effects stemming from the band structure of the finite graphene layer. 

The asymmetry of the bands with respect to the wavevector is better explained in \fref{fig:dos_valley} using chiral $p$-wave symmetry. The results in \fref{fig:dos} and \fref{fig:tres} are calculated for one of the valleys. 
For the other valley, the pair potential is also given by \eref{eq:pairing} but with the change $q\!\rightarrow\!-q$. As a consequence, the asymmetric FPR bands inside the Dirac cone are reflected in the $q$-axis for the other valley. The symmetry with respect to the energy in the DOS is thus recovered when the contribution from both valleys is considered together, see \fref{fig:dos_valley}(c). The change of valley does not affect the dispersion relation of the IBS and SABS for chiral symmetries, and chirality is preserved in the total contribution of the spectral density, as it is shown in \fref{fig:dos_valley}. 

The finite size is manifested by the appearance of discrete bands instead of a continuous spectrum like in \fref{fig:IBS}. For the undoped cases with $E_F\!=\!0$, e.g., \fref{fig:dos}(a-d), a band appears inside the gap at the Dirac point. A second band can be perceived close to the gap edge, for the symmetries with a full gap around $q\!\sim\!0$, like $d_{x^2-y^2}$ in \fref{fig:dos}(a) and $p_x$ in \fref{fig:tres}(a). 
To better analyze the FPR, we consider a heavily-doped regime with $E_{F}\!=\!10\Delta$ in \fref{fig:dos}(e-h) and \fref{fig:tres}(e-h). The extra bands emerging from high doping appear as wavy arc-shaped bands framed by the anisotropic superconducting gap and no ZZES band is present as predicted by \eref{eq:FP}. 

We now focus on the effect of proximity induced unconventional pairing in graphene. 
For the $d_{x^2-y^2}$-wave symmetry, graphene's band structure is deformed according to the $\cos2\theta$ dependence of the pairing amplitude, cf. \fref{fig:dos}(a). As a result, the DOS features a $V$-shaped gapped profile, even in the presence of FPR, see \fref{fig:dos}(e). 
Even after adding up all the momentum channels in the DOS, we can still observe in the left panel of \fref{fig:dos}(a) a small contribution from the layer's second band as a small peak below the gap edge. 
For $d_{xy}$-wave symmetry, where the gap edge now follows a $\sin2\theta$ dependence, there is a clear zero-energy peak in the DOS coming from the emergence of a flat band in the spectrum, see \fref{fig:dos}(d,h). 
The intermediate instance between these symmetries is well represented by $r_1\!=\!0.4$, \fref{fig:dos}(c,g), where the flat band acquires a dispersion at the same time that the gap edge is deformed similarly to the $d_{xy}$-wave case. The DOS captures such a superposition of $d$-wave states displaying a smaller gap with increased DOS but still featuring a minimum at zero energy. The FPR can now mask this effect in the DOS, cf. \fref{fig:dos}(g), but the local minimum at $E\!=\!0$ remains. 
The situation where both $d_{x^2-y^2}$ and $d_{xy}$ weight exactly the same in the pairing states is the chiral $d$-wave ($\chi_{d}$) case, shown in \fref{fig:dos}(b,f). For this chiral symmetry the effect of the SABS is better perceived: the DOS is finite but features a $U$-shaped gap profile with sharp edges and the SABS crossing the gap is clearly visible in the spectral density. 

For $p$-wave symmetries, we find analogous results with some important differences. 
Similarly to the $d_{xy}$-wave case, $p_x$-wave symmetry features a zero-energy peak in the DOS from a flat band, independently of the doping level, see \fref{fig:tres}(a,e). 
Analogously, $p_y$-wave symmetry features a $V$-shaped DOS comparable to that of $d_{x^2-y^2}$-wave, as shown in \fref{fig:tres}(d,h). It is important to notice that for $p_y$-wave there are no resonances at $E\!=\!\pm\Delta$, a characteristic feature of $p$-wave superconductors. 
In the presence of disorder, $d_{xy}$ and $p_x$-wave (and, correspondingly, $d_{x^2-y^2}$ and $p_y$-wave) display different behavior and can be thus distinguished better\cite{Lu_2016,Burset_2017}. 

The chiral $p$-wave symmetry, $\chi_p$, shows an interesting difference with respect to the $\chi_d$ case. The $\chi_p$ SABS have a linear dispersion which results in a convex enhanced DOS below the gap. Interestingly, there is still a minimum at $E\!=\!0$, stemming from graphene's band structure. Indeed, the convex enhanced DOS is suppressed around $|E|\!\sim\!0$ in \fref{fig:tres}(b,c) for the undoped case. For the doped situation, the linear chiral SABS has two different contributions. Outside graphene's band, it mixes with the IBS and features a linear dispersion responsible for the enhanced DOS. Close to zero energy, however, the SABS mixes with a FPR resonance that always crosses zero at $\hbar q\!\sim\!0$. 
Additionally, the wavy FPR bands become small peaks in the DOS for subgap energies. 

The doped case reveals an important difference between the $d$- and $p$-wave symmetries. Comparing \fref{fig:dos}(e,f,g) and \fref{fig:tres}(e,f,g), we immediately observe that, for the same set of system parameters, the $d$-wave symmetry cases, with the notable exception of $r_1\!=\!0$, feature an even number of FPR bands while the number is odd for $p$-wave symmetry. Additionally, one of the $p$-wave bands is always zero at $\hbar q\!\sim\!0$. 
This is a consequence of the different symmetry classification  in two dimensions of the topologically trivial $d$-wave case, when $r_1\!\neq\!0$ in \eref{eq:pairing}, and the nontrivial $p$-wave and $d_{xy}$-wave cases\cite{Schnyder_2008,Sato_2009,Sato_2010,Schnyder_2010,Budich_2013,Kashiwaya_2014}. 

\begin{figure*}[tbp]
	\includegraphics[width=1.\textwidth]{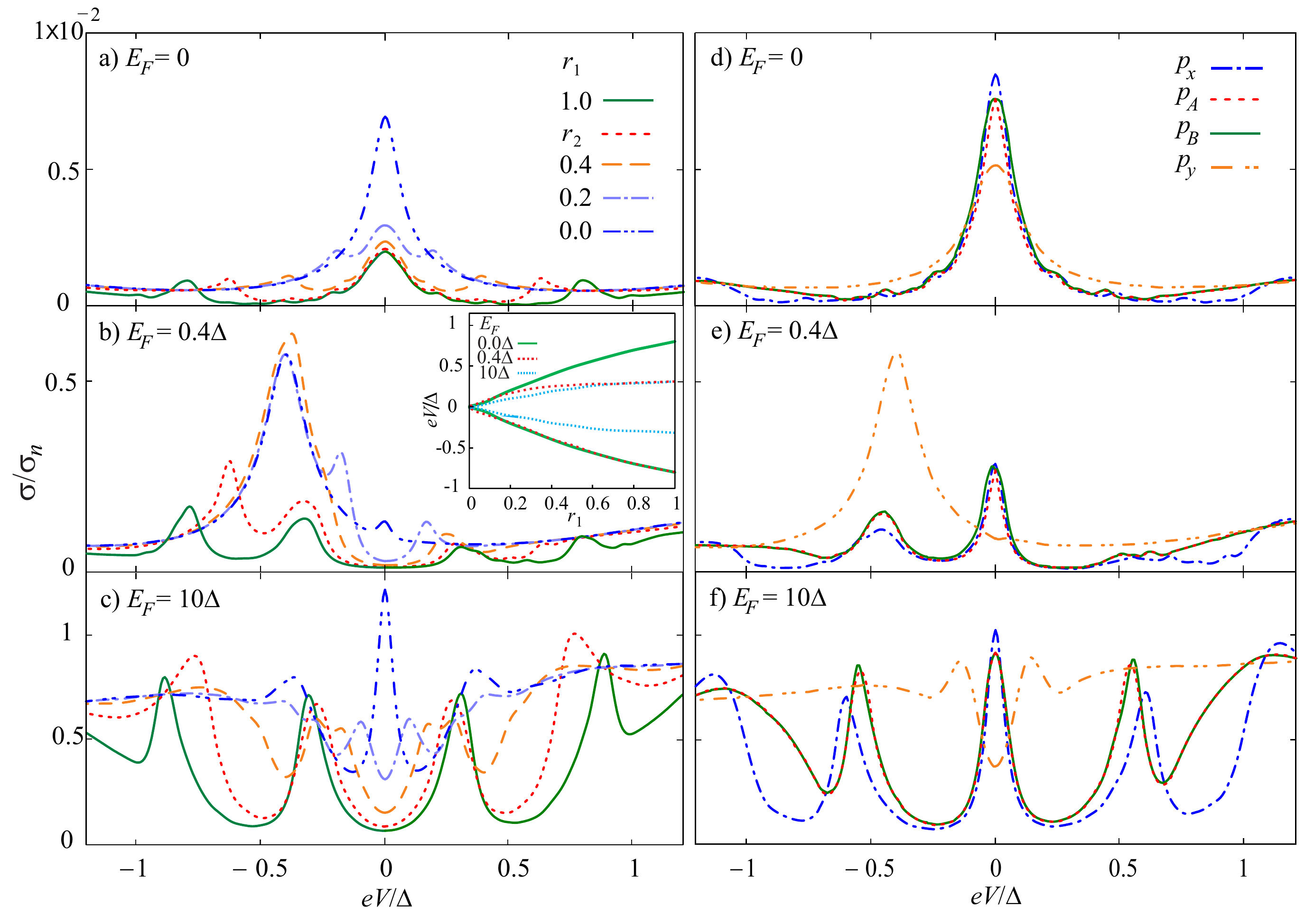}
	\caption{Differential conductance of the graphene N-G-S
		junction. 
		(a,b,c) Conductance for $d$-wave superconductors with different values of the pairing phase. The inset in (b) shows the dependence of the first resonance on $r_{1}$ for different values of the Fermi energy of the graphene stripe. 
		(d,e,f) Similar results for $p$-wave superconductors. The Fermi energy of the graphene layer is taken at $E_{F}\!=\!0$ (a,d), $E_{F}\!=\!0.4\Delta $ (b,e), and $E_{F}\!=\!10\Delta $ (c,f). In all cases, $E_{FN}\!=\!E_{FS}\!=\!30\Delta $, $t_{L}\!=\!0.1$, $t_{R}\!=\!1$, and $d\!=\!2\protect\xi $. }
	\label{fig:cond}
\end{figure*}

\begin{figure*}[tbp]
	\includegraphics[width=1.\textwidth]{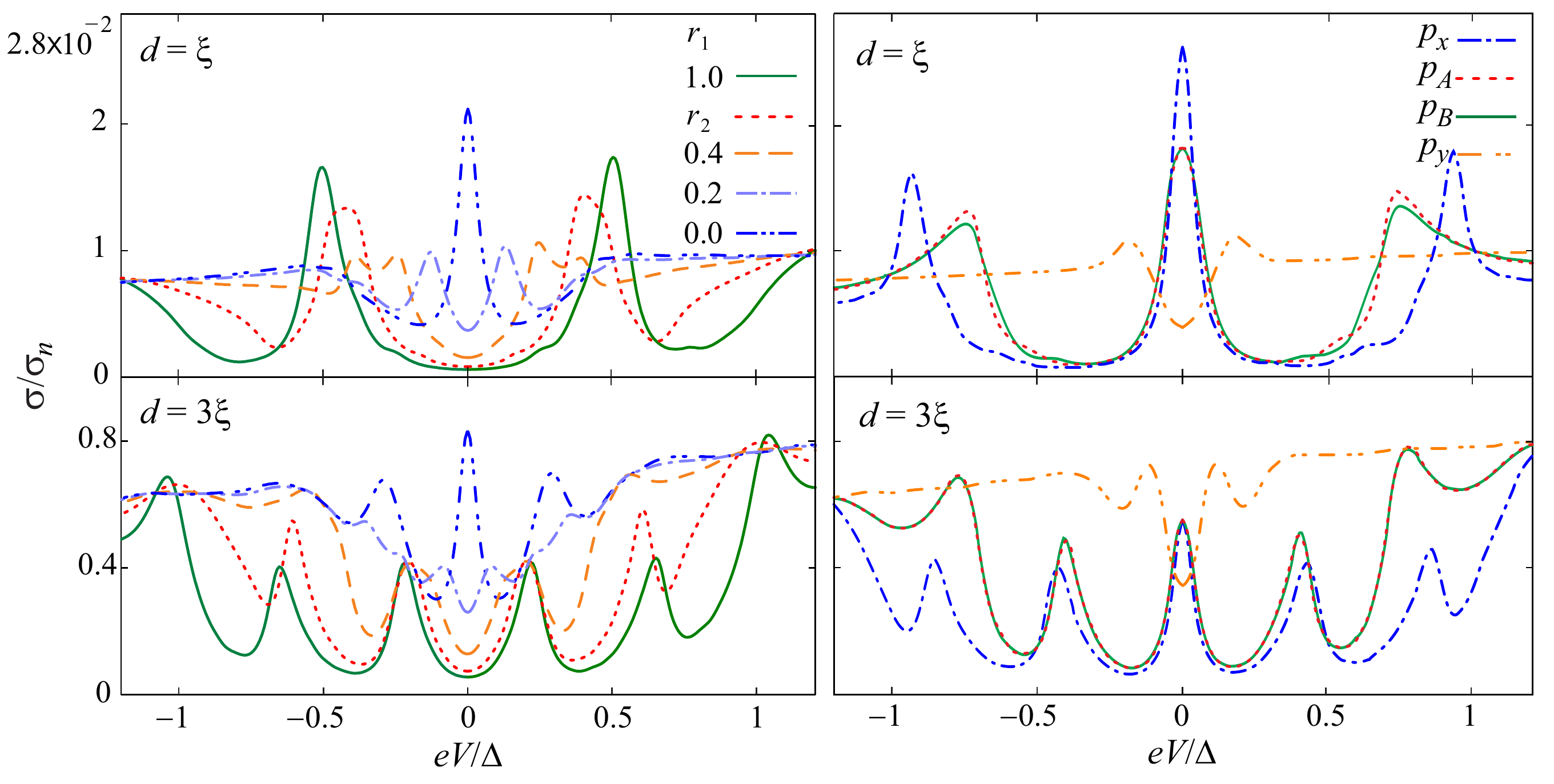}
	\caption{Differential conductance of the graphene N-G-S
		junction for different widths. 
		(Left) Conductance for $d$-wave superconductors with different values of the pairing phase. 
		(Right) Similar results for $p$-wave superconductors. The graphene layer's width is $d\!=\!\xi$ (top) and $d\!=\!3\xi$ (bottom). In all cases, $E_{FN}\!=\!E_{FS}\!=\!30\Delta $, $E_{F}\!=\!10\Delta $, $t_{L}\!=\!0.1$, and $t_{R}\!=\!1$. }
	\label{fig:extra}
\end{figure*}


\section{Differential conductance \label{sec:cond}}
We now analyze the differential conductance in the N-G-S junction sketched in \fref{fig:sketch}(a). 
To this end, we plot \eref{eq:cond} in \fref{fig:cond} for $d$-wave (left column) and $p$-wave symmetries (right column). Results for $s$-wave are similar to those of $d_{x^{2}-y^{2}}$. 

We start analyzing $s$- and $d$-wave symmetries. In \fref{fig:cond}(a), we plot the conductance at $E_{F}\!=\!0$ for different values of $r_1$, see \fref{fig:sketch}(c). In all cases, there is a strong zero-bias conductance peak (ZBCP). When $r_{1}\!\neq \!0$, we also observe two small peaks at the position of the effective superconducting gap. The ZBCP in this setup is mostly due to the contribution of graphene's ZZES at the N-G interface, where $t_{L}\!=\!0.1$. However, for the case with $r_{1}\!=\!0$ where the superconductor features a nodal, flat band, the ZBCP is greatly enhanced since now merges the SABS with the ZZES. This is a signature of the gap closing and edge state for $d_{xy}$-wave pairing. As $r_{1}$ increases from zero, and the $d_{xy}$-wave mixes with the $d_{x^{2}-y^{2}}$-wave, this zero-energy state splits giving rise to the effective gap edge, cf. \fref{fig:dos}(a). We show the evolution of the split conductance peaks from finite $r_{1}$ in the inset of \fref{fig:cond}(b).

By setting $E_{F}\!\neq \!0$, as it is done in \fref{fig:cond}(b,c), the ZZES moves away from zero voltage. In these doped cases, a ZBCP only appears for $d_{xy}$-wave symmetry ($r_1\!=\!0$). As we increase the doping of the central graphene region, the FPR become more pronounced, see \fref{fig:cond}(b,c). The main reason is that the Andreev processes at the G-S interface become retro-reflections for $E_{F}\!>\!\Delta $, which favors the formation of closed trajectories in the G region. The geometric origin of the FPR is more clearly shown in \fref{fig:extra}, where we plot the conductance with the same parameters as in \fref{fig:cond}(c) but with $L\!=\!\xi $ (top) and $L\!=\!3\xi $ (bottom).

We now consider $p$-wave pairing symmetries. As in the previous section, we only consider $p_{x}$-, $p_{y}$-, and chiral $p$-wave symmetries. For the undoped graphene layer with $E_{F}\!=\!0$, shown in \fref{fig:cond}(d), the conductance features a clear ZBCP. For $p_{y}$-wave pairing, the ZBCP is mostly due to the ZZES and has the smallest value because of the absence of any SABS band, cf. \fref{fig:tres}(d). For the other symmetries, both the ZZES and the SABS contribute to the ZBCP. The highest value of the peak corresponds to $p_{x}$-wave state, where the SABS-IBS is a flat nodal surface state, as it is shown in \fref{fig:tres}(a). 

Setting $E_{F}\!\neq \!0$ in \fref{fig:cond}(e,f), we clearly see that the ZBCP survives for all symmetries except for the trivial $p_{y}$-wave. In the strongly doped case with $E_{F}\!=\!10\Delta$, the ZBCP coexists with the FPRs for the nontrivial cases. Comparing \fref{fig:cond}(c) and \fref{fig:cond}(f), calculated with $L\!=\!2\xi$, it is clear that the $d$-wave cases (with $r_1\!\neq\!0$) feature an even number of peaks, while the $p_x$- and chiral $p$-wave cases have the additional ZBCP. The number of resonances is determined by the length of the graphene layer as shown in \fref{fig:extra}. 

It is interesting to note that the magnitude of the ZZES peak at $eV\!=\!-E_{F}$ in relation to the ZBCP seems to follow the opossite behavior for $p$- and $d$-wave cases, cf. \fref{fig:cond}(b) and \fref{fig:cond}(e). 
The ZZES peak in \fref{fig:cond}(e) is rather small for the nontrivial $p_{x}$- and chiral $p$-wave states, but has a quite pronounced contribution in the $p_{y}$-wave case. 
In contrast, a strong ZZES peak appears when nontrivial $d_{xy}$-wave symmetry becomes dominant but it is weak for the $d_{x^2-y^2}$-wave dominated cases. 
The magnitude of the ZZES peak is completely determined by the ratio between the real and imaginary parts of the pairing as defined in \eref{eq:pairing}. 
When the real part dominates, i.e., for $r_1\!\rightarrow\!1$ ($p_x$- and $d_{x^2-y^2}$-waves), the effective gap edge $\Delta_\text{eff}$ is maximum around $\hbar q\!\sim\!0$  ($\Delta_\text{eff}\approx\left\vert \Delta\left( \theta=0 \right) \right\vert$). 
On the contrary, when the imaginary part becomes dominant for $r_1\!\rightarrow\!0$ ($p_y$- and $d_{xy}$-waves), the effective gap edge around $\hbar q\!\sim\!0$ is reduced, merging with the ZZES when $E_{F}\!\lesssim \!\Delta _{\mathrm{eff}}$. 
The dependence of the effective gap with $\hbar q$ for the different pairings is clearly observed in \fref{fig:dos}(a,d) and \fref{fig:tres}(a,d). 
The ZZES appears for values of $\hbar q$ that are close to zero for $d_{xy}$- and $p_y$-wave ($r_1\!\rightarrow\!0$) and far away from zero for $d_{x^2-y^2}$- and $p_x$-wave ($r_1\!\rightarrow\!1$). 
After averaging over the incident modes, the ZZES thus provides a stronger contribution around $|\hbar q|\!\sim\!0$ for pairings with $r_1\!\rightarrow\!0$, resulting in the strong peaks at $eV\!=\!-E_{F}$ in \fref{fig:cond}(b) and \fref{fig:cond}(e). 

Finally, for the strongly doped case in \fref{fig:cond}(f) we clearly observe some of the characteristic behaviors of $p$-wave superconductors. Chiral $p$- and $p_{x}$-wave states feature a clear ZBCP due to the ZES bands showed in \fref{fig:tres}(f,g). The $p_{y}$-wave pairing displays a $V$-shape gap, with a finite minimum even though the conductance is calculated in the tunnel limit.

\section{Conclusion \label{sec:conc}}
Motivated by recent experimental advances in the implementation of graphene--superconductor ballistic junctions\cite{Klapwijk_2015,Geim_2016,Efetov_2016}, we have theoretically studied the transport properties of a ballistic, finite-size graphene layer contacted by a normal and a superconducting lead. We particularly considered the emergence of unconventional superconductivity in the graphene layer\cite{Robinson_2017,Jarillo_2018}. 

Using a microscopic description based on Green's functions techniques, we included in our model several experimentally relevant issues like the Fabry-P\'erot resonances originated by the finite length of the graphene layer, the different transmission of the graphene-reservoir interfaces, and the presence of graphene's edge states. 
We calculated the spectral density, DOS and differential conductance of the graphene junction in the presence of unconventional superconductivity with different forms of $d$- or $p$-wave symmetry. 
We find that for energies below the gap, both the DOS and conductance show a very intricate profile due to the presence of several types of resonances. 
In addition to the Fabry-P\'erot resonances and graphene's edge states, we identify the emergence of Andreev surface states and interface bound states with different dispersions. 
Our analytical results allows us to identify the separate contribution from each state to the DOS and their impact on the differential conductance. We thus determine the optimal conditions for the detection of unconventional
superconductivity using graphene layers.

In particular, we find that the presence of graphene's zigzag edge states can mask the emergence of a ZBCP if the superconducting pairing allows for one. A finite doping is enough to separate and distinguish the contributions from ZZES and SABS to the conductance. 
In the presence of high doping compared to the superconducting gap, the geometrical FPRs become stronger. However, the subgap SABS from the induced unconventional pairing still have clear signatures in the spectral density and the DOS. Such resonances do not hide the ZBCP originating from $d_{xy}$-, $p_{x}$- or chiral $p$-wave states, for lengths of the graphene layer comparable to the superconducting coherence length. 
Additionally, these nontrivial pairings always display an odd number of conductance resonances. Even in the presence of high doping, the FPRs mix with the SABS but the topological zero energy states are still present in the spectral density and result in additional zero bias peaks in the conductance. 

Our results provide a useful guide for future experiments that combine graphene with unconventional superconductors or that study emerging unconventional superconductivity in graphene induced by the asymmetry of the hexagonal lattice. 


\acknowledgments 
We acknowledge funding from COLCIENCIAS, project No. 110165843163 and doctorate Scholarship 617, the European Union's Horizon 2020 research and innovation programme under the Marie Sk\l odowska-Curie Grant No. 743884, and Spanish MINECO through Grant No.~FIS2014-55486-P and through the ``Mar\'{\i}a de Maeztu'' Programme for Units of Excellence in R\&D (MDM-2014-0377).



\appendix

\section{Boguliubov-de Gennes-Dirac Hamiltonian. \label{sec:app1}}
We consider a semi-infinite graphene layer with a zigzag edge along the $y$-axis at $x_{L}\!=\!0$ and extending into the $x\!>\!0$ half-plane.
With this orientation the Brillouin zone has Dirac points (valleys) at $\mathbf{K}_{\pm }=(0,\pm K)$. The conserved momentum along the $y$ direction is $\hbar q$. The wave function for the sublattice $A(B)$ is then given by
\begin{equation*}
\Psi _{A(B)}\left( \mathbf{r}\right) =\e^{iKy}\Phi _{A(B)}^{+}\left( \mathbf{r%
}\right) +\e^{-iKy}\Phi _{A(B)}^{-}\left( \mathbf{r}\right) ,
\end{equation*}
where the functions $\Phi _{A(B)}^{\pm }\left( \mathbf{r}\right) $
are solutions of a 2D Dirac equation
\begin{align*}
\hat{H}_{\pm }={}& \hat{p}_{x}\sigma _{x}\pm q\sigma _{y}, \\
0={}&\left( \hat{H}_{\pm }-\left( E_{F}+E\right) \right) \Phi ^{\pm }\left( \mathbf{r} \right) , \\
\Phi ^{\pm }\left( \mathbf{r}\right) ={}&\left( 
\begin{array}{cc}
\Phi _{A}^{\pm }\left( \mathbf{r}\right) & \Phi _{B}^{\pm }\left( \mathbf{r}%
\right)%
\end{array}%
\right) ^{T} \\ ={}& \left( 
\begin{array}{cc}
f_{A}^{\pm }\left( x\right) & f_{B}^{\pm }\left( x\right)%
\end{array}%
\right) ^{T}\e^{iqy}.
\end{align*}
Zigzag edges are formed by a line of atoms of only one of graphene's sublattices (A or B) and do not mix valleys. If we adopt a Dirichlet boundary condition, we get 
\begin{align*}
\Psi _{A(B)}\left( \mathbf{r}\right)={}&\e^{iKy}\Phi _{A(B)}^{+}\left(
x_{L},y\right) + \e^{-iKy}\Phi _{A(B)}^{-}\left( x_{L},y\right) =0, \\
\Rightarrow &f_{A(B)}^{\pm }\left( x_{L}\right) =0.
\end{align*}
Then, we can consider the boundary problem separately and use only one valley. The Dirac-Bogoliubov-de Gennes (DBdG) Hamiltonian for a 2D graphene sheet adopts the form
\begin{equation*}
H_{BDG}=\left( 
\begin{array}{cc}
\check{H}-E_{F}\check{I} & \check{\Delta}(\mathbf{k})\check{I} \\ 
\check{\Delta}^{\dag }(\mathbf{k})\check{I} & E_{F}\check{I}-T\check{H}T^{-1}%
\end{array}%
\right) ,
\end{equation*}
where we are considering the weak-coupling approximation ($\mathbf{k}$ fixed on the Fermi surface) where the order parameter is only angle dependent\cite{Linder_2008}, i.e.,
\begin{equation*}
\check{\Delta}(\mathbf{k})=\check{\Delta}(\theta )\text{,}
\end{equation*}
and $\check{H}$ is the single-particle Hamiltonian in sublattice and valley spaces
\begin{equation*}
\check{H}=\left( 
\begin{array}{cc}
\hat{H}_{+} & 0 \\ 
0 & \hat{H}_{-}%
\end{array}%
\right) .
\end{equation*}

Valley degeneracy allows us to consider only one of the two valley sets. Then, by using that $\left[ \check{H},T\right] \!=\!0$, the $8\times8$ matrix decouples to a $4\times 4$ matrix equation for zigzag edges, namely, 
\begin{equation}
H_{BDG}=\left( 
\begin{array}{cc}
\hat{H}_{\pm }-E_{F}\sigma _{0} & \Delta (\mathbf{k})\sigma _{0} \\ 
\Delta ^{\dag }(\mathbf{k})\sigma _{0} & E_{F}\sigma _{0}-\hat{H}_{\pm }%
\end{array}%
\right) .
\end{equation}
We adopt for the pair potential the following anisotropic symmetries [see \fref{fig:sketch}(c)]
\begin{align*}
\Delta (\theta_{\pm} )={}&\Delta _{1}\cos (n\theta_{\pm}) +i\Delta
_{2}\sin (n\theta_{\pm}) , \\
n ={}&0\text{ \ for }s\text{-wave,}  \notag \\
n ={}&1\text{ \ for }p\text{-wave,}  \notag \\
n ={}&2\text{ \ for }d\text{-wave,}  \notag
\end{align*}
where the parameters $\Delta _{1}$ and $\Delta _{2}$ obey the relations
\begin{align*}
\Delta _{i} \equiv &r_{i}\Delta_{0}\text{,} \\
\sqrt{r_{1}^{2}+r_{2}^{2}} =&1,
\end{align*}
with $i\!=\!1,2$, and their respective phases are defined by
\begin{equation*}
\varphi _{\pm }=-i\ln \left( \frac{\Delta\left( \theta _{\pm }\right) }{%
\left\vert \Delta \left( \theta _{\pm }\right) \right\vert }\right) \text{.}
\end{equation*}

\section{Green's function of graphene layer with zigzag edges and induced unconventional superconductivity. \label{sec:app2}}
The solutions of the DBdG equations have the form
\begin{align*}
\psi _{\pm }^{e} ={}& \e^{\pm ik_{e}x}\left( 
u_{0}\phi _{e}^{\pm } , v_{0} \e^{-i\varphi _{\pm }}\phi _{e}^{\pm }%
\right) ^{T}\text{,} \\
\psi _{\pm }^{h} ={}& \e^{\pm ik_{h}x}\left( 
v_{0}\phi _{h}^{\pm } , u_{0} \e^{-i\varphi _{\pm }}\phi _{h}^{\pm }%
\right) ^{T}\text{,}  \notag
\end{align*}
with
\begin{align*}
\phi _{e\left( h\right) }^{+} ={}&\left( 
1 , \e^{i\alpha _{e\left( h\right) }}%
\right)^T , \, \phi _{e\left( h\right) }^{-}=\left( 
1 , - \e^{-i\alpha _{e\left( h\right) }}%
\right)^T ,  \label{exp} \\
\e^{i\alpha _{e(h)}} ={}&\hbar v_F\frac{k_{e(h)}+iq}{E_{F}\pm E} \, , \quad \Omega =\sqrt{E^{2}-\left\vert \Delta \right\vert ^{2}},  \notag \\
k_{e\left( h\right) } ={}&\sgn\left( E_{F}\pm \Omega \right) \sqrt{\frac{%
\left( E_{F}\pm \Omega \right) ^{2}}{\hbar ^{2}v_F^{2}}-q^{2}},  \notag \\
u_{0} ={}&\sqrt{\frac{1}{2}\left( 1+\frac{\Omega }{E}\right) } , \, v_{0}=\sqrt{\frac{1}{2}\left( 1-\frac{\Omega }{E}\right) } . \notag 
\end{align*}
The wavefunctions $\psi _{\pm }^{\varepsilon }$ propagate under a pair potential $\Delta $, while their conjugates,  $\bar{\psi}_{\pm }^{\varepsilon }$, move under $\Delta^{\ast }$. 
Therefore the functions $\bar{\psi}_{\pm}^{\varepsilon }$ can be constructed from the solutions $\psi_{\pm }^{\varepsilon }$ by changing $\Delta $ by $\Delta ^{\ast }$ and multiplying by the conjugation matrix $\sz$ (see more details in Ref.~\onlinecite{Herrera_2010}). 
For a semi-infinite system with one edge, the asymptotic solutions of the DBdG equations are a superposition of normal reflection and Andreev reflections as follows
\begin{align*}
\psi _{<}^{e} =&\psi _{-}^{e}+r_{e}\psi _{+}^{e}+r_{h}\psi _{-}^{h}\text{,}
\\
\psi _{<}^{h} =&\psi _{+}^{h}+r_{h}^{\prime }\psi _{-}^{h}+r_{e}^{\prime
}\psi _{+}^{e}\text{,}  \notag \\
\psi _{>}^{e} =&\psi _{+}^{e}\text{,}  \notag \\
\psi _{>}^{h} =&\psi _{-}^{h}\text{,}  \notag
\end{align*}
where $r_{e(h)}$ are the reflection coefficients. As a boundary condition we adopted a zigzag border of atoms of sublattice B, so that the B component must be zero at $x\!=\!0$. It then follows that
\begin{align*}
r_{e} =& \e^{-2i\alpha _{e}}\frac{\left( 1-\Gamma _{0}^{2}\right) }{1-\Gamma
_{0}^{2} \e^{-i\Delta \varphi }}=r_{h}^{\prime } \e^{i\Delta \varphi} \e^{-2i\alpha _{h}}\e^{-2i\alpha _{e}}, \\
r_{h} =&\frac{\Gamma _{0}\left( \e^{-i\Delta \varphi }-1\right) }{1-\Gamma
_{0}^{2}\e^{-i\Delta \varphi }}\left( \e^{i\alpha _{h}}\e^{-i\alpha
_{e}}\right) =r_{e}^{\prime },  \notag \\
\Gamma _{0} =&\frac{v_{0}}{u_{0}}.  \notag
\end{align*}

From the asymptotic solutions that obey specific boundary conditions at the left ($<$) and right ($>$) edges of a ribbon, we construct the Green's function as\cite{McMillan_1968,Furusaki_1991,Kashiwaya_2000,Herrera_2010,Burset_2015,Crepin_2015,Breunig_2018}
\begin{equation}
\check{g}_{q}(x,x^{\prime })=\left\{\!\begin{array}{cc}
\sum\limits_{\mu,\nu}\check{C}_{\mu\nu }\psi
_{<}^{\mu }\left(q,x\right) \bar{\psi}_{>}^{\nu T}\left(q,x^{\prime }\right)
\check{\gamma} ,& x>x^{\prime } \\
\sum\limits_{\mu,\nu}\check{C}_{\mu \nu }^{\prime
}\psi _{>}^{\mu }\left(q,x\right) \bar{\psi}_{<}^{\nu T}\left(q, x^{\prime
}\right) \check{\gamma} ,& x<x^{\prime } 
\end{array} \right. ,
\end{equation}
where $\mu,\nu\!=\!e,h$ label electron- and hole-like solutions of the DBdG equations and we include $\check{\gamma}\!=\!\ti\sz$, with $\ti$ the identity matrix in Nambu space, to ensure covariance\cite{Herrera_2010}. 
By integrating \eref{eq:Green_equation} on the
infinitesimal interval $(x^{\prime }-\epsilon ,$ $x^{\prime }+\epsilon )$, with $\epsilon\!\ll\!1$, we
obtain the continuity relation
\begin{equation}
\lim_{\epsilon \rightarrow 0}\left[ \check{g}_{q}\left( x^{\prime }+\epsilon
,x^{\prime }\right) -\check{g}_{q}\left( x^{\prime }-\epsilon ,x^{\prime
}\right) \right] =-\frac{i}{\hbar v_F}\tz\sx,
\label{eq:continuity_relation}
\end{equation}
with the Pauli matrix $\tz$ acting in Nambu space. 
From \eref{eq:continuity_relation} it is possible to determine the coefficients $\check{C}_{\mu \nu}^{(\prime )}$. 

From the continuity relation we deduce the matrix coefficients
\begin{align*}
\check{C}_{eh\left( he\right) }^{\left( \prime \right) } ={}&0,  \notag \\
\check{C}_{ee} ={}&\check{C}_{ee}^{\prime }=\check{C}_{hh}=\check{C}%
_{hh}^{\prime }=\frac{i}{\hbar v_F}\frac{1}{FH-AB}\check{Y}, \\
\check{Y} ={}&\left( 
\begin{array}{cccc}
-B & 0 & H & 0 \\ 
0 & -B & 0 & F \e^{i\varphi _{-}} \e^{i\varphi _{+}} \\ 
-F & 0 & A & 0 \\ 
0 & -H \e^{-i\varphi _{+}} \e^{-i\varphi _{-}} & 0 & A%
\end{array}%
\right) ,  \notag \\
A ={}&\Gamma _{0}^{2}\left( \e^{i\alpha _{h}}+ \e^{-i\alpha _{h}}\right) -\left(
\e^{i\alpha _{e}}+ \e^{-i\alpha _{e}}\right) , \\
B ={}&\left( \e^{-i\alpha _{h}}+ \e^{i\alpha _{h}}\right) -\Gamma _{0}^{2}\left(
\e^{i\alpha _{e}}+ \e^{-i\alpha _{e}}\right) , \\
H ={}&\Gamma _{0}\left[ \e^{-i\varphi _{-}}\left( \e^{-i\alpha
_{h}}- \e^{-i\alpha _{e}}\right) \right. \notag \\ & \left. + \e^{-i\varphi _{+}}\left( \e^{i\alpha
_{h}}- \e^{i\alpha _{e}}\right) \right] \e^{i\varphi _{-}} \e^{i\varphi _{+}}, \\
F ={}&\Gamma _{0}\left[ \e^{-i\varphi _{+}}\left( \e^{-i\alpha
_{h}} - \e^{-i\alpha _{e}}\right) \right. \notag \\ & \left. + \e^{-i\varphi _{-}}\left( \e^{i\alpha
_{h}}- \e^{i\alpha _{e}}\right) \right] .
\end{align*}
Thus, the Green function for $x<x^{\prime }$ is given by the expression
\begin{align*}
\check{g}_{R}(x,x^{\prime }) ={}&\check{C}_{ee} \left[ \psi _{-}^{e}\left(x\right) \bar{\psi}_{+}^{eT}\left( x^{\prime }\right) +r_{e}\psi_{+}^{e}\left( x\right) \bar{\psi}_{+}^{eT}\left( x^{\prime }\right) \right. 
\\
& \left. +r_{h}\psi _{-}^{h}\left( x\right) \bar{\psi}_{+}^{eT}\left( x^{\prime}\right) +\psi _{+}^{h}\left( x\right) \bar{\psi}_{-}^{hT}\left( x^{\prime}\right) \right. \\
& \left. +r_{h}^{\prime }\psi _{-}^{h}\left( x\right) \bar{\psi}_{-}^{hT}\left(x^{\prime }\right) +r_{e}^{\prime }\psi _{+}^{e}\left( x\right) \bar{\psi}_{-}^{hT}\left( x^{\prime }\right) \right] .
\end{align*} 
where the $q$ dependence has been omited. A similar expression is obtained for $x\!>\!x^{\prime }$ by exchanging the signs of the subindexes and changing $e,h\rightarrow h,e$ in the superindexes. For the particular case with $\left( x,x^{\prime }\right) \!=\!\left(
0^{+},0\right) $, the Green function is given by
\begin{equation}
\check{g}_{R}=\frac{i}{\hbar v_F}\left( 
\begin{array}{cccc}
\frac{\Psi _{31}H-B\Psi _{11}}{FH-AB} & 0 & \frac{\Psi _{33}H-B\Psi _{13}}{%
FH-AB} & 0 \\ 
-1 & 0 & 0 & 0 \\ 
\frac{A\Psi _{31}-F\Psi _{11}}{FH-AB} & 0 & \frac{A\Psi _{33}-F\Psi _{13}}{%
FH-AB} & 0 \\ 
0 & 0 & 1 & 0%
\end{array}%
\right) ,
\end{equation}
with
\begin{align*}
\Psi _{11} ={}& 1+\Gamma _{0}^{2}+2\Gamma _{0}r_{h}+r_{e}\left( \Gamma
_{0}^{2} \e^{2i\alpha _{h}} \e^{2i\alpha _{e}} \e^{-i\Delta \varphi }+1\right) , \\
\Psi _{31} ={}& \Gamma _{0}\left( \e^{-i\varphi _{-}}+ \e^{-i\varphi _{+}}\right)
+r_{h}\left( \e^{-i\varphi _{-}}+\Gamma _{0}^{2} \e^{-i\varphi _{+}}\right) \\
&+\Gamma _{0} \e^{-i\varphi _{+}}r_{e}\left( \e^{2i\alpha _{h}} \e^{2i\alpha
_{e}}+1\right) , \\
\Psi _{13} ={}& \Gamma _{0}\left( \e^{i\varphi _{-}}+ \e^{i\varphi _{+}}\right)
+r_{h}\left( \e^{i\varphi _{+}}+\Gamma _{0}^{2} \e^{i\varphi _{-}}\right) \\
&+\Gamma _{0} \e^{i\varphi _{-}}r_{e}\left( \e^{2i\alpha _{h}} \e^{2i\alpha
_{e}}+1\right) , \\
\Psi _{33} ={}& 1+\Gamma _{0}^{2}+2\Gamma _{0}r_{h}+r_{e}\left( \e^{2i\alpha
_{h}} \e^{2i\alpha _{e}}+\Gamma _{0}^{2} \e^{-i\Delta \varphi }\right) .
\end{align*}

Following the same procedure, the Green function of a normal graphene stripe of length $d$ is given by
\begin{equation*}
\check{g}_{0}(x,x^{\prime }>x)=\frac{-i}{\hbar v_F}\left( 
\begin{array}{cc}
\hat{g}_{e} & 0 \\ 
0 & \hat{g}_{h}%
\end{array}%
\right) ,
\end{equation*}
with
\begin{align*}
\hat{g}_{e\left( h\right) } =&\frac{ \e^{\pm i\left( x^{\prime }-x\right)
k_{e\left( h\right) }}}{D_{e\left( h\right) }}\left( 
\begin{array}{cc}
IK & \mp s \e^{\pm i\alpha }IL \\ 
\mp s \e^{\mp i\alpha }JK & JL%
\end{array}%
\right) _{e(h)}\text{,} \\
D_{e\left( h\right) } =&\left( \e^{-i\alpha _{e\left( h\right) }}+ \e^{i\alpha
_{e\left( h\right) }}\right) \left( 1+ \e^{\mp 2i\alpha _{e\left( h\right)
}} \e^{\pm 2idk_{e\left( h\right) }}\right) \text{,} \\
I_{e\left( h\right) } =&1+ \e^{\mp 2i\alpha _{e\left( h\right) }} \e^{\pm
2idk_{e\left( h\right) }} \e^{\pm 2ixk_{e\left( h\right) }}\text{,} \\
J_{e\left( h\right) } =&1- \e^{\pm 2idk_{e\left( h\right) }} \e^{\pm 2ixk_{e\left( h\right) }}\text{,} \\
K_{e\left( h\right) } =&1- \e^{\mp 2ix^{\prime }k_{e\left( h\right) }}\text{,}
\\
L_{e\left( h\right) } =&1+ \e^{\mp 2i\alpha _{e\left( h\right) }} \e^{\mp
2ix^{\prime }k_{e\left( h\right) }}\text{,}
\end{align*}
and $s\!=\!1$. 
For $x\!>\!x^{\prime }$ the Green function is obtained from the transpose of the last expression by interchanging the coordinates ($x\!\leftrightarrow\! x^{\prime }$) and setting $s\!=\!-1$. For a semi-infinite graphene sheet, we have $D_{e\left( h\right) }\!=\!e^{-i\alpha _{e\left( h\right)}} + \e^{i\alpha _{e\left( h\right) }}$ and $F_{e\left( h\right) }\!=\!J_{e\left(h\right) }\!=\!1$. Since Green functions for graphene with zigzag edges depend on the order of the spatial arguments, the following convention was adopted for Dyson's equation [\eref{eq:Dyson_equation}] thats couples two regions with edges at $x\!=\!0$, namely, 
\begin{align*}
\check{G}_{ij}(x,x^{\prime }) ={}& \check{g}_{ij}(x,x^{\prime })+\check{g}_{iR}(x,0^{-})\check{\Sigma}_{RL}%
\check{G}_{Lj}(-0^{+},x^{\prime }) \\
={}& \check{g}_{ij}(x,x^{\prime }) + \check{g}_{iR}(x,-0^{-})\check{\Sigma}_{LR}%
\check{G}_{Rj}(0^{+},x^{\prime }) ,
\end{align*}
where $0^{\pm }$ are positive infinitesimal real numbers
satisfying $0^{-}\!<\!0^{+}$. 
For example for $\check{G}_{LL}$ we obtain
\begin{align*}
\check{G}_{LL} ={}&\check{g}_{LL}\left( I+\Sigma _{LR}M_{RR}\check{g}%
_{RR}\Sigma _{RL}\check{g}_{LL}\right) \text{,} \\
M_{RR}={}&\left[ I-\check{g}_{RR}\Sigma _{RL}\check{g}_{LL}\Sigma _{LR}\right]
^{-1}\text{.} 
\end{align*}
For the model of a highly doped graphene superconductor electrode ($\e^{i\alpha _{e\left( h\right) }}\!=\!1$) coupled with $t\!=\!1$ to a graphene film of length $d$ the last Green function has the following denominator
\begin{equation}
D =\left( 1- \e^{-i\Delta \varphi }\Gamma _{0}^{2}\right) ^{2}X ,
\end{equation}
with
\begin{align}
X ={}&1+\frac{\Gamma _{0}^{2}+ \e^{i\Delta \varphi /2}}{\Gamma
_{0}^{2}- \e^{i\Delta \varphi /2}}\left( \e^{-i\alpha _{+}}h_{+} + \e^{i\alpha
_{-}}h_{-}\right) \notag \\ & + \e^{-i\alpha _{+}} \e^{i\alpha _{-}}h_{+}h_{-} .  \notag 
\end{align}
The first factor contains the SABS dispersion relation in \eref{eq:SABS}, namely, 
\begin{gather*}
1-\e^{-i\Delta \varphi }\Gamma _{0}^{2}=\left( E^{2}-E_{SABS}^{2}\right)
/\Lambda \text{,} \\
\Lambda =\frac{1}{2}\e^{i\Delta \varphi /2}\left( E+\Omega \right) \left(
\Omega \cos \left( \Delta \varphi /2\right) -iE\sin \left( \Delta \varphi
/2\right) \right) ,
\end{gather*}
where $\Lambda $ is responsible for some effects of superconducting phase chirality in the SABS. The factor $X$ encodes the IBS and FPR dispersion relations [\eref{eq:IBS-FP}], 
\begin{gather*}
X\Lambda _{X}=\left( E^{2}-E_{IBS-FPR}^{2}\right) /\Lambda _{X}, \\
\Lambda _{X}=\frac{E\left( \Phi -1\right) -\left( 1+\Phi \right) \Omega }{%
\left( \e^{-i\alpha _{+}}h_{+}+ \e^{i\alpha _{-}}h_{-}\right) \left(
1-Z^{2}\right) } \\ \times
\frac{E\left[ C\left( \Phi -1\right) +\left( 1+\Phi \right) \right] -\Omega %
\left[ \left( 1-\Phi \right) -C\left( 1+\Phi \right) \right] }{\left(
C\left( \Phi -1\right) +\left( 1+\Phi \right) \right) ^{2}}\text{,}
\end{gather*}
with $\Phi \!=\! \e^{-i\Delta \varphi }$. 
Here $\Lambda_{X}$ also includes some effects of valley and superconducting phase chirality.

\bibliography{graphene_NGS}

\end{document}